\documentclass[a4paper,fleqn,usenatbib]{mnras}

\usepackage{newtxtext,newtxmath}

\usepackage[T1]{fontenc}
\usepackage{ae,aecompl}

\usepackage{graphicx}	
\usepackage{amsmath}	
\usepackage{amssymb}	
\usepackage{bm}		       

\title[Overdense Region of LAEs at z=6.5]{GTC Observations of an Overdense Region of LAEs at z=6.5}

\author[K. Chanchaiworawit \& R. Guzm\'{a}n et al.]{
K. Chanchaiworawit,$^{1}$\thanks{E-mail: krittapas@ufl.edu}
R. Guzm\'{a}n,$^{1,2}$
J.M. Rodr\'{i}guez Espinosa,$^{3,4}$
 \newauthor
 N. Castro Rodr\'{i}guez,$^{3,4}$
E. Salvador-Sol\'{e},$^{2}$
R. Calvi,$^{3,4}$ 
J. Gallego,$^{5}$
\newauthor
A. Herrero,$^{3,4}$
A. Manrique,$^{2}$
A. Mar\'{i}n Franch,$^{6}$
J.M. Mas-Hesse,$^{7}$
\newauthor
I. Aretxaga,$^{8}$
E. Carrasco,$^{8}$ 
E. Terlevich,$^{8}$
and R. Terlevich$^{8,9}$
\\
$^{1}$Department of Astronomy, University of Florida, 211 Bryant Space Science Center, Gainesville, FL, 32611, USA\\
$^{2}$Institut de Ciencies del Cosmos. Universitat de Barcelona, UB-IEEC. Mart\'{i} Franqu\'{e}s 1, E-08028 Barcelona, Spain\\\
$^{3}$Instituto de Astrof\'{i}sica de Canarias, E-38205 La Laguna, Spain\\
$^{4}$Departamento de Astrof\'{i}sica, Universidad de La Laguna, E-38205 La Laguna, Spain\\
$^{5}$Departamento de Astrof\'{i}sica y CC de la Atm\'{o}sfera, Universidad Complutense de Madrid, Spain\\
$^{6}$CEFCA, Plaza san Juan 1, E-44001 Teruel, Spain\\
$^{7}$Centro de Astrobiolog\'{i}a - Dept. de Astof\'{i}sica (CSIC-INTA), Madrid, Spain\\
$^{8}$Instituto Nacional de Astrof\'{i}sica Optica y Electr\'{o}nica, AP 51 y 216, 72000, Puebla, Mexico\\
$^{9}$Institute of Astronomy, University of Cambridge, Madingley Road, Cambridge CB3 0HA, UK
}

\date{Accepted XXX. Received YYY; in original form ZZZ}

\pubyear{2017}

\begin{document}
\label{firstpage}
\pagerange{\pageref{firstpage}--\pageref{lastpage}}
\maketitle

\begin{abstract}

	We present the results of our search for the faint galaxies near the end of the Reionisation Epoch. This has been done using very deep OSIRIS images obtained at the Gran Telescopio Canarias (GTC). Our observations focus around two close, massive Lyman Alpha Emitters (LAEs) at redshift 6.5, discovered in the SXDS field within a large-scale overdense region ~\citep{ouchi2010}. The total GTC observing time in three medium band filters (F883w35, F913w25 and F941w33) is over 34 hours covering $7.0\times8.5$ arcmin$^2$ (or $\sim30,000$ Mpc$^3$ at $z=6.5$). In addition to the two spectroscopically confirmed LAEs in the field, we have identified 45 other LAE candidates. The preliminary luminosity function derived from our observations, assuming a spectroscopic confirmation success rate of $\frac{2}{3}$ as in previous surveys, suggests this area is about 2 times denser than the general field galaxy population at $z=6.5$. If confirmed spectroscopically, our results will imply the discovery of one of the earliest protoclusters in the universe, which will evolve to resemble the most massive galaxy clusters today.
	
\end{abstract}

\begin{keywords}
Reionisation -- Large-Scale Structure of Universe -- Early Universe -- Observations -- High-Redshift -- Distances and Redshifts
\end{keywords}



\section{Introduction}



	Observation of high redshift galaxies and galaxy clusters provides a basic information about the large scale structure formation of the universe. The higher the redshifts of the galaxies, the further we look back in time. Thus, observing the galaxies as far back as the time of their assembly is ideal for studying the early evolution of the Universe. However, detecting these high redshift galaxies is challenging due to their low surface brightness. Nevertheless, novel observation techniques combined with large telescopes and their instruments allow us to detect many high-z galaxies, especially Lyman $\alpha$ (Ly$\alpha$) Emitters (LAEs) and Lyman Break Galaxies (LBGs) (e.g.,~\citet{hu1998},~\citet{rhoads2000},~\citet{kudritzki2000},~\citet{steidel1996, steidel1999}, ~\citet{bouwens2010, bouwens2014, bouwens2015},~\citet{ellis2013},~\citet{giavalisco2004},~\citet{laporte2014},~\citet{mclure2010},~\citet{ouchi2003, ouchi2008, ouchi2010},~\citet{shapley2003}, and~\citet{taniguchi2005}).
	
	LAEs and LBGs are star forming galaxies with strong Ly$\alpha$ emission, the latter possessing significantly stronger UV-continuum than the former~\citep{haiman1999}. These galaxies have long since been theorised to be observable up to the Reionisation Epoch~\citep{meier1976, partridge1967}. Studying these galaxies is crucial to understanding the complete picture of the total reionisation of the intergalactic medium (IGM) in the early universe. The luminosity functions of high-z LAEs and LBGs are established to be steeper than the low-z populations~\citep{bouwens2015, finkelstein2015}. This leads to the conclusion that the majority of the ionising photons responsible for reionisation of the intergalactic neutral hydrogen is produced from the young stellar populations of low-mass star forming galaxies (e.g.~\citet{erb2015},~\citet{dressler2011, dressler2015},~\citet{henry2012}, and~\citet{yan2010}). 
	
	Additionally, one way to constrain the cosmological parameters, particularly the matter and dark energy density parameters ($\Omega_{M}$ and $\Omega_\Lambda$), is to trace the large scale structures formed around regions with enhanced dark matter density. Thus, studying the evolution of the galaxy clusters mass function is key to understanding the influence of dark matter and dark energy to the history of the universe (e.g.~\citet{allen2011},~\citet{demianski2015},~\citet{gonzalez2015}, and~\citet{vikhlinin2009}). Both galaxy clusters and groups from low to intermediate redshift ($z<1$) have been studied extensively (e.g.,~\citet{carlberg2001},~\citet{eke2004},~\citet{blakeslee2003},~\citet{ellis1997},~\citet{halliday2004},~\citet{holden2005},~\citet{homeier2005}, and~\citet{stanford1998}). High-z galaxy clusters in the process of assembling (protoclusters) are sometimes discovered fortuitously in galaxy surveys (e.g.~\citet{ouchi2005, ouchi2008},~\citet{shimasaku2003}, and~\citet{steidel1998}), or via direct protocluster searches around massive sources, as they tend to be signposts of high matter concentrations (e.g.~\citet{barr2004},~\citet{debreuck2002, debreuck2003},~\citet{leferve1996},~\citet{overzier2006b, overzier2009},~\citet{reuland2004},~\citet{sanchez1999, sanchez2002},~\citet{venemans2007}, and~\citet{zheng2006}). 

	Near the end of the Reionisation Epoch is the first ideal observing window for the large scale structure formation and galaxy clusters assembly. Even though intrinsically fainter than quasars, LAEs are more suitable for probing the faint end of high redshift star forming galaxy luminosity function (e.g.~\citet{kashikawa2011}). Moreover, the visibilities of LAEs are enhanced, though marginally, when they are in groups or clusters, due to the ionised cavity in the IGM ~\citep{dayal2009, dayal2011, hutter2015, miralda1998, mortlock2011}. Therefore, to search for the highest redshift large scale structure formation, we conduct a survey for a protocluster at $z=6.5$, right before the end of the Reionisation Epoch, using massive LAEs as signposts of high matter concentration. 

	We have selected the part of Subaru/XMM-Newton Deep Survey or SXDS field~\citep{furusawa2008}, which exhibits a sign of an overdensity by containing 2 spectroscopically confirmed massive LAEs at redshift $\sim6.5$ discovered by~\citet{ouchi2008, ouchi2010}. The 2 LAEs are only $\sim$300 kpc apart, assuming they are relaxed, and yield star formation rates between $\sim25-45\:M_{\odot}/yr$ based on the estimation by~\citet{ouchi2010}. The typical dark matter halo mass for these massive LAEs is $\sim2\times10^{11}\:M_{\odot}$ (e.g.~\citet{gawiser2007},~\citet{kovac2007}, and~\citet{ouchi2010, sobacchi2015}). We have conducted a photometric selection for the LAE candidates at z=6.5 from this field centred around the 2 massive LAEs. Our goals are 1) to photometrically select LAE candidates at z=6.5 down to a flux limit fainter than in previous studies; and 2) to determine the level of overdensity and sign of protocluster in this sub-field by comparing the LAE luminosity functions. 

	The clustering properties and the expected final mass of such galaxy cluster at $z=0$ produced by the observed overdensity are discussed in detail by~\citet{jmre2016} (paper-II). The result of this work will help to connect how massive galaxy clusters assemble and collapse as seen from the local to high redshift Universe, such as the massive galaxy cluster at $z\sim1.19$ discovered by~\citet{gonzalez2015}. The magnitudes presented in this work are given in AB system~\citep{okegunn1983}. Throughout this paper, we have adopted $\Lambda CDM$ cosmology with $\Omega_{\Lambda} = 0.7$, $\Omega_{M} = 0.3$, and $h = 0.7$.

\section{Observations and Data Reduction}
\subsection{Observation Strategies}
		The observations were carried out during Semester 2011B and 2012B on the 10.4-meter Gran Telescopio Canarias (GTC) on the summit of the Canary Island, La Palma, Spain. We utilised the Optical System for Imaging and low-to-intermediate-Resolution Integrated Spectroscopy (OSIRIS) in imaging mode. We applied our 3-band photometry and dropout criteria in selecting LAE candidates, rather than the traditional narrow-band search (e.g.,~\citet{pritchet1987, pritchet1990},~\citet{cowie1988},~\citet{Djorgovski1992},~\citet{Djorgovski1993},~\citet{Macchetto1993},~\citet{thompson1995},~\citet{shimasaku2006},~\citet{ouchi2008, ouchi2010},~\citet{matthee2015}, and~\citet{penin2015}). Our 3-band photometry used the 3 reddest intermediate-band (15-35 nm) filters from the SHARDS program, which studies red and dead galaxies through their absorption features, covering a contiguous spectral window between 5000 to 9500 $\AA$~\citep{pablo2013}. 
		
	 	In order to prove whether the field has both an overdensity and evidence of a protocluster, we aim to reach a sensitivity down to $F_{Ly\alpha} \geq  5\times10^{-18}\:erg\:s^{-1}\:cm^{-2}$. This guarantees that we can complete the LAE Luminosity Function (LF) down to $log(L_{Ly\alpha})=42.4\:erg\:s^{-1}$. To reach this level of sensitivity, we observe through the medium-band filters: F883w35, F913w25, and F941w33 (henceforth, F883, F913, and F941), with total exposure time in each filter of 12.25, 10.78, and 11.30 hours, respectively. The central pointing of the field is  R.A. = 02:18:20.350 (hh:mm:ss) and Dec = -04:34:28.80 (dd:mm:ss). We utilise a 6-point dithering pattern, tracing a parallelogram with 8 arcsec base and 16 arcsec height. The exposure times per frame in F883, F913, and F941 were 350s, 400s, and 300s, respectively. The details of the observations are also shown in Table~\ref{tab:1}. The median seeing at  $\sim$9000 $\AA$ and the median airmass during the observing runs were 0.7'' and 1.20, respectively.

\subsection{Data Reduction Processes}		

				In order to obtain the final reduced images, we conduct the following image reduction routines for the science images in all bands. First, we obtain the master flat and bias frames for each band by calculating the pixel-to-pixel median bias and dome-flat values. All the science frames are bias corrected and flat fielded in a standard manner. The near-infrared night sky is mainly dominated by the glow of OH and $O_2$ emission lines in the atmosphere~\citep{osterbrock1996, rousselot2000}. 
We subtract the sky background from the science frames by creating a master sky image for each band. To do so, we use Source Extractor (SExtractor, hereafter)~\citep{ba1996} through the de-biased, flat-fielded science frames to create mask images (objects' flux profiles and positions) of the detected sources in each band. Sky frames are created by subtracting the extracted sources from the science frames. However, these sky frames contain holes from the masking process. We patch up these holes through the following process. We define 4 sub-areas, 25$\times$25 pixel$^2$ boxes, locating 25 pixels to the left, right, up, and down from the pixel that needs to be patched (reference pixel). The 25-pixel clearance from the box is set to ensure that all the pixels in the boxes are the actual sky background and not part of extended objects or saturated stars. Next, we obtain the median and standard deviation of the sky background in each box. The calculated median and standard deviation of the sky background at the reference pixel are the average of the up-down and left-right interpolation results. The re-assigned sky value for each masked pixel is taken from a random value of a gaussian distribution characterised by the calculated median and standard deviation. Then, the master sky image for each band is obtained from the pixel-to-pixel median of the patched sky frames in each filter. 
				
				Before sky subtraction, we scale the master sky image to the same sky level on each science frame. The scaling factors are calculated through the following steps. We calculate the median sky value in each of five 200$\times$200 pixel$^2$ boxes, located on the four corners and the centre of each science frame. The large sub-areas are assigned to ensure that the median sky background values calculated within the boxes would not be dominated by any object's flux. Then, we calculate the ratios between those median values and the ones from the corresponding regions in the master sky image. The scaling factor is the median of all the 5 ratios. After that, we subtract the scaled master sky images from all corresponding science frames. 
				
 				We use IRAF~\citep{iraf1, iraf2} to shift, align, and combine (median) all science frames for each filter of the first and second chip of OSIRIS, separately. Then, we use IRAF to combine the images from the two chips together and trim out the part of the final reduced images with sufficiently low signal-to-noise, particularly the left edge ($\sim$0.5-arcmin wide) of the first OSIRIS chip. The shifts used in stacking of science frames in our case are decimal point of a pixel shifts with linear interpolation in resampling and stacking to minimise the effect of cosmic rays. However, to address concerns that the linear interpolation may not efficiently account for the noise level in the final images, we have conducted a simulation on F913 frames to compare 2 different interpolants: i.e., linear and cubic-spline. We have found that in the case of final F913 image with linear interpolation, the RMS noise levels are $9.70 \pm 0.5$ ADUs and $12.2\pm 0.7$ ADUs in intermediate and high noise regions, respectively. While, in the case of final F913 image with cubic-spline interpolation, the RMS noise levels are $10.0 \pm 0.8$ ADUs and $13.0\pm 1.0$ ADUs in intermediate and high noise regions, respectively. For our situation, both interpolation methods yield similar results within the uncertainty. Thus, the linear interpolation in resampling and stacking of science frames, which we have utilised, is sufficient.
 				
 				Even though the observations were carried out in the best weather conditions (dark sky, low cloud coverage, low vapour, and typical seeing of 0.7 arcsec), the photometric calibration for zero-point magnitude in each band using the standard stars, such as G158-100, G191-B2B, G24-9, Feige34, Feige110, and Ross640, has also been done. The 3$\sigma$ limiting AB magnitudes for the final reduced F883, F913, and F941 images are 26.54, 26.56, and 25.84, respectively. The quoted 3$\sigma$ limiting AB magnitudes are the SExtractor's auto-magnitudes. Similarly, the 3$\sigma$ limiting aperture magnitudes are the SExtractor's aperture magnitudes, which produce a signal-to-noise ratio of 3 within a 2-arcsec aperture. The limiting aperture magnitudes of the F883, F913, and F941 images are 26.87, 26.80, and 26.30, respectively. 
		
\begin{table*}
	\centering
	\caption{Details of the observing runs and the quality of the final reduced images in all 3 bands.}
	\label{tab:1}
	\begin{tabular}{cccccccc} 
		\hline
		\hline
		Band & 	Central Wavelength        &	FWHM	  & Exposure   &       PSF  &        Area  		  &   mag$_{lim}$                 &          	                               \\
		         &                  ($\AA$)       &        ($\AA$)         & 		(s)		     &     (arcsec)     &          (arcmin$^2$)   &                                         &                       Date(s) of Observations        \\
	       (1)     & 			(2)		          & 		(3)     &	      (4)                & 	(5)   	  &			(6)		  & 		(7)			          &           	     		      	\\
	        \hline
	           & 				     & 		  &	           & 		  &			  & 			          &           	     		      	\\	        		  
	           & 				     & 		  &	           & 		  &			  & 			          &           	     		    	\\	        
		  F883w35 &              8800           &        340         & 		44100		     &     0.83     &         59.37 (63.75)   &          26.54 (26.87)           & 2012 Sept 11-14, 21, 25; Nov 18;        \\	
	           & 				     & 		  &	           & 		  &			  & 			                    	     		                         &     Dec 7, and 16-17; 2013 Jan 14-17	\\	        		  
	           & 				     & 		  &	           & 		  &			  & 			          &           	     		      	\\	        		  	           
		  F913w25 &              9100          &        280         & 		38800		      &     0.80     &         59.37 (63.75)   &          26.56  (26.80)          &                       2012 Sept 14-17, 19, 25        \\	        
	           & 				     & 		  &	           & 		  &			  & 			          &           	     		      	\\	        
	           & 				     & 		  &	           & 		  &			  & 			          &           	     		      	\\	        
		  F941w33 &              9410          &        340         & 		40700	   	       &     0.82     &         59.37 (63.75)   &          25.84 (26.28)            &                       2011 Oct 26-28; 2012 Dec 20;       \\	        		  		                   
		   & 				           & 		  &	           & 		  &			  & 			        			                                              &           	     2013 Jan 2, 4  		      	\\	  
	           & 				    & 		  &	           & 		  &			  & 			          &           	     		      	\\	        		        	          
			\hline
	\end{tabular}
	
	\raggedright
	\textbf{Notes.} (1) Filter name; (2) filter's central wavelength in $\AA$; (3) filter's FWHM in $\AA$; (4) total exposure time of each band in seconds; (5) FWHM of an unsaturated star; (6) total survey areas in arcmin$^2$, after (before) trimming; (7) 3$\sigma$ limiting magnitudes, auto- (2-arcsec aperture) magnitudes.
\end{table*}






\section{Analysis}


\subsection{Sources Extraction}

		Final F883, F913, and F941 reduced images are aligned, shifted, and trimmed to have the same size and pixel-by-pixel coordinates. First, we run SExtractor on each image individually. To push the detections as deep as possible, we set the detection threshold to be 0.85$\times$ RMS above the median background. The minimum area for detection is set to 9 pixel$^2$, the minimum contrast for de-blending to 10$^{-6}$, and the spurious cleaning efficiency to 5.0 (1.0 for maximum cleaning efficiency, and 10.0 for no cleaning). Zero-point magnitudes are 32.68, 32.54, and 32.12 for the F883, F913, and F941 images, respectively. 
		
\begin{figure}
	\includegraphics[width=\columnwidth]{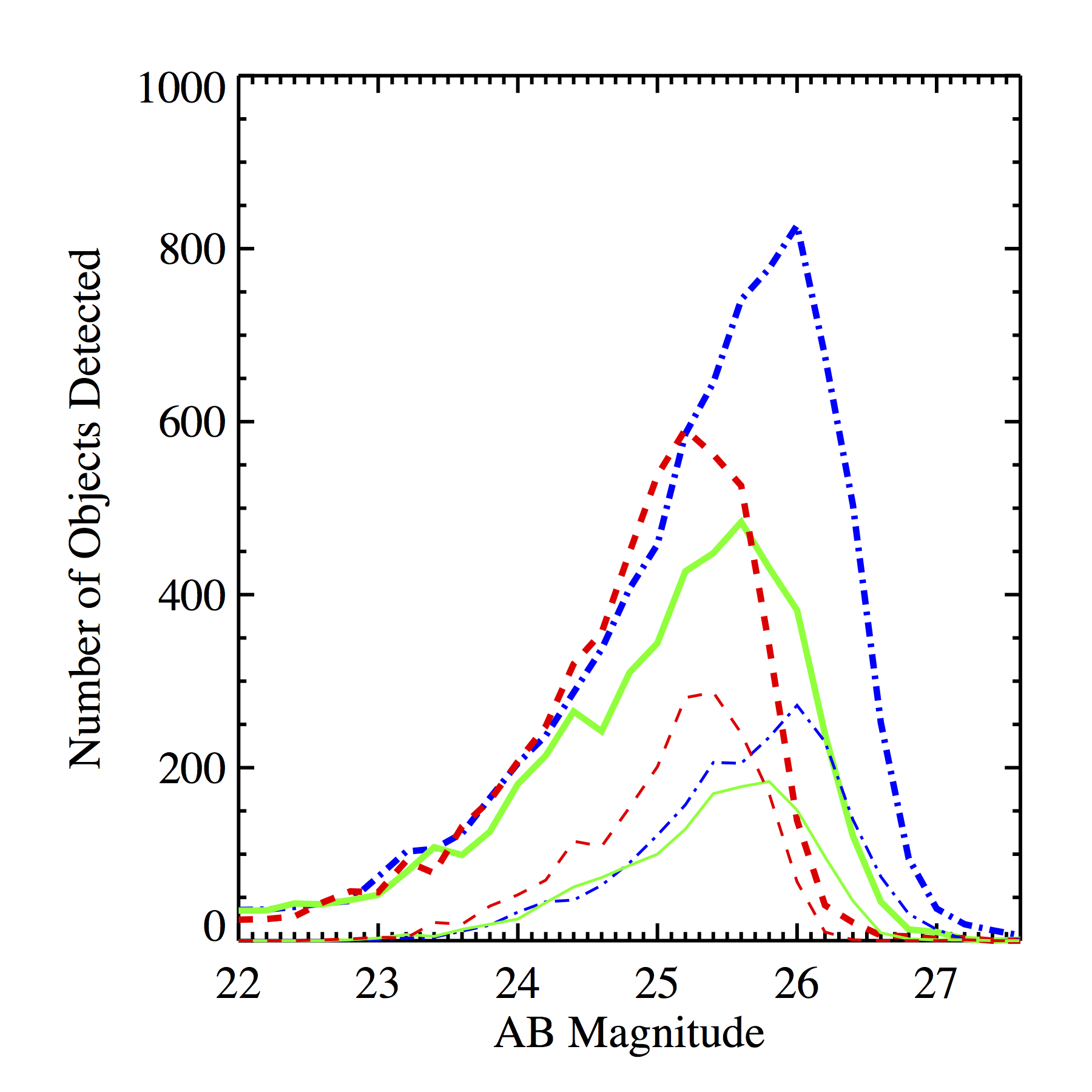}
    \caption{Magnitude histogram of objects detected in each band, with the bin size of 0.2 mag. The thicker blue dash-dotted, green solid, and red dashed lines represent the numbers of real detections in the F883, F913, and F941 bands, respectively. The thinner blue dash-dotted, green solid, and red dashed lines represent the numbers of the spurious detections from the negative F883, F913, and F941 images, respectively.}
    \label{fig:1}
\end{figure}

		For the purpose of extracting sources with accurate flux measurements, we run SExtractor twice on each image. The first run of SExtractor yields the RMS noise background map and sky background for each image. These 2 images are used for creating a weighting map for each band. The weighting maps are used in the second run of SExtractor for treating the differential gains in different regions of the images. The second SExtractor run on each image also provides a mask image. The mask images consist of flux profiles and positions of the objects without any background. The histograms of sources detected in all 3 images (F883, F913, and F941) by SExtractor, using the mentioned SExtractor parameters, are shown in Figure~\ref{fig:1}. These preliminary extractions are performed for the purpose of image quality assessment, and not for the selection of LAE candidates. 
		
		Next, we compute contamination levels of all bands. We use the mask images created from SExtractor to zero-out the fluxes of the objects in each image. Then, we create the negative images from the masked science images, by multiplying the arrays by $-1$. The objects in the negative frames are the local minima associated with the noise spikes, fringes patterns, and dithering pattern holes (spurious sources) in the science frames. Then, we run the SExtractor on the negative images with the same set of parameters as used for the real detection. Statistically, the objects detected in negative frames represent the spurious detections in the real science images. Nevertheless, dithering holes only appear in the negative images and should not be included as the sources of contamination in the positive images. Thus, we use the dithering pattern as a map to identify these negative objects and exclude them from the spurious sources. Then, by calculating the ratio between the numbers of spurious and real detections in each band for all magnitude bins, we can determine the levels of contamination.
		
\begin{figure}
	\includegraphics[width=\columnwidth]{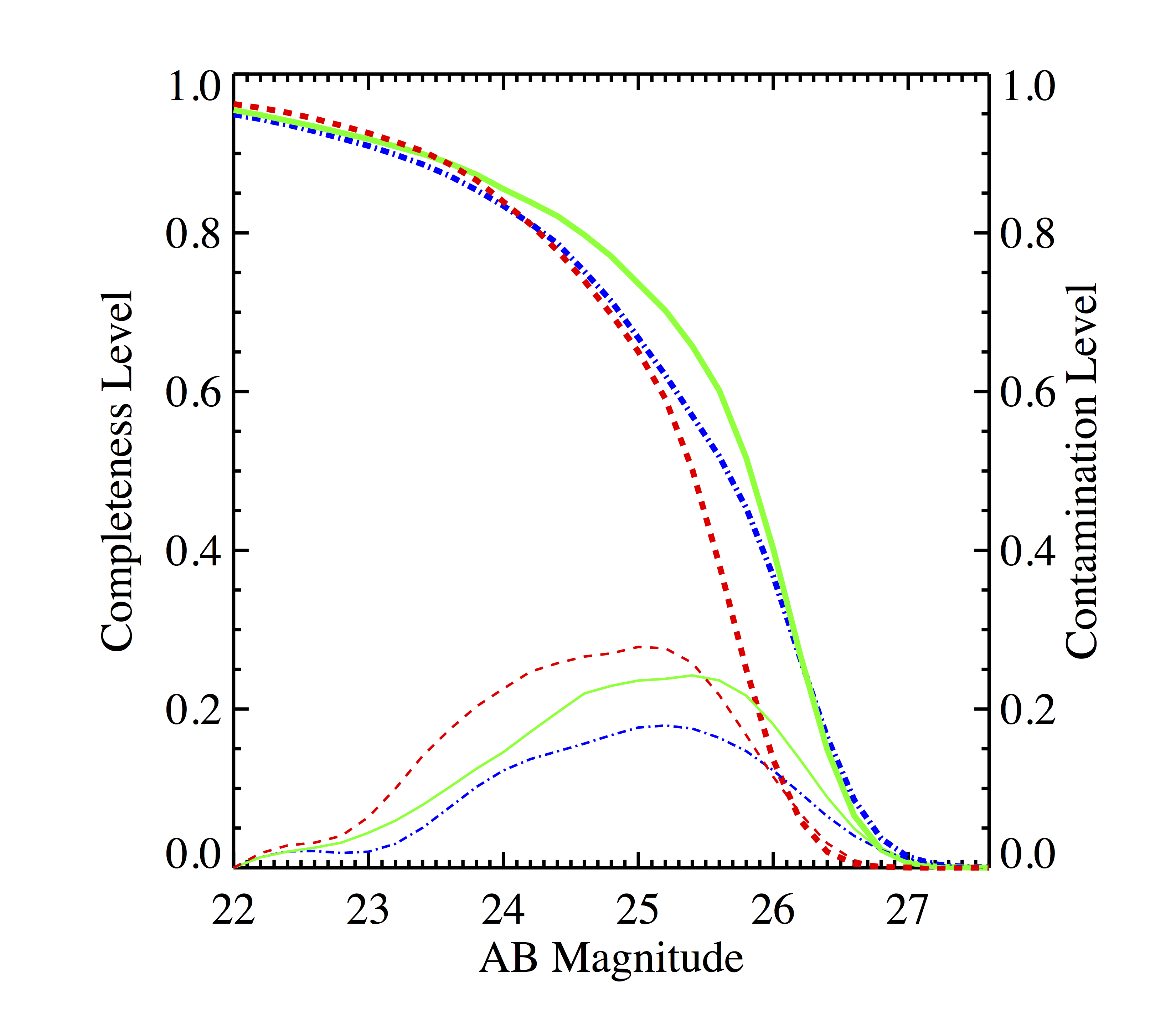}
    \caption{Completeness and contamination levels as function of AB magnitude for all bands. The thicker blue dash-dotted, green solid, and red dashed lines represent completeness levels of the F883, F913, and F941bands, respectively.
While, the corresponding thinner lines represent contamination levels.
}
    \label{fig:2}
\end{figure}

		Then, the completeness levels are obtained in the following manner. We use IRAF to generate objects with compact, almost star-like flux profiles ($"$stars$"$). The simulated stars are injected into each image randomly, with uniform magnitude and spatial distribution (2000 stars/frame, ranging from 22 to 28 magnitude). The compactness and flux profile of a simulated star are adequate to mimic the appearance of LAEs at high redshift. The completeness levels are derived from the ratio between the numbers of objects recovered by SExtractor at the same positions of the injected stars with deviations in magnitude within $\pm$0.5 mag. These contamination and completeness levels are used for depth and quality assessment, and are shown as functions of AB magnitude (SExtractor's auto-magnitude) in Figure~\ref{fig:2}.

		Figure~\ref{fig:1} reveals that the F883 image contains the largest number of objects, both from real and spurious detections. While, the F913 image contains the smallest number of objects of both types. The estimated total number of real detections in F883, F913, and F941 images are 9000, 5000, and 7000, respectively; while, the estimated number of spurious detections are 2000, 1500, and 2400, respectively. The level of spurious detections in the F941 image is the highest among all bands for magnitude brighter than 25.4 mag. This suggests that F941 is prone to spurious sources and lack of depth. The levels of completeness and spurious contamination in all 3 bands shown in  Figure~\ref{fig:2} reflect the relative quality of the images. The quality of the F913 image is the best among all bands, both in terms of depth (completeness level $=$50\% at 25.8 mag) and noise level (contamination level $<$20\% at 25.8 mag), whereas the quality of the F941 image is the worst among all bands, both in terms of depth (completeness level $=$50\% at 25.4 mag) and noise level (maximum contamination level $\sim$27\% at 25.4 mag). For F941, the peak number of sources detected is at 25.2 mag, 0.4 and 0.8 mag brighter than the F913 and F883 images, respectively. As stated in Table~\ref{tab:1}, the 3$\sigma$ limiting magnitude of the F941 image is the brightest among all bands. However, at the 3$\sigma$ limiting magnitude, the completeness levels are 20\%; while, the contamination levels are about 15\% for all 3 bands. These differences in image quality affect our selection criteria for the LAE candidates, which will be discussed in detail in the next section. 
		
\begin{figure}
	\includegraphics[width=1.0\columnwidth]{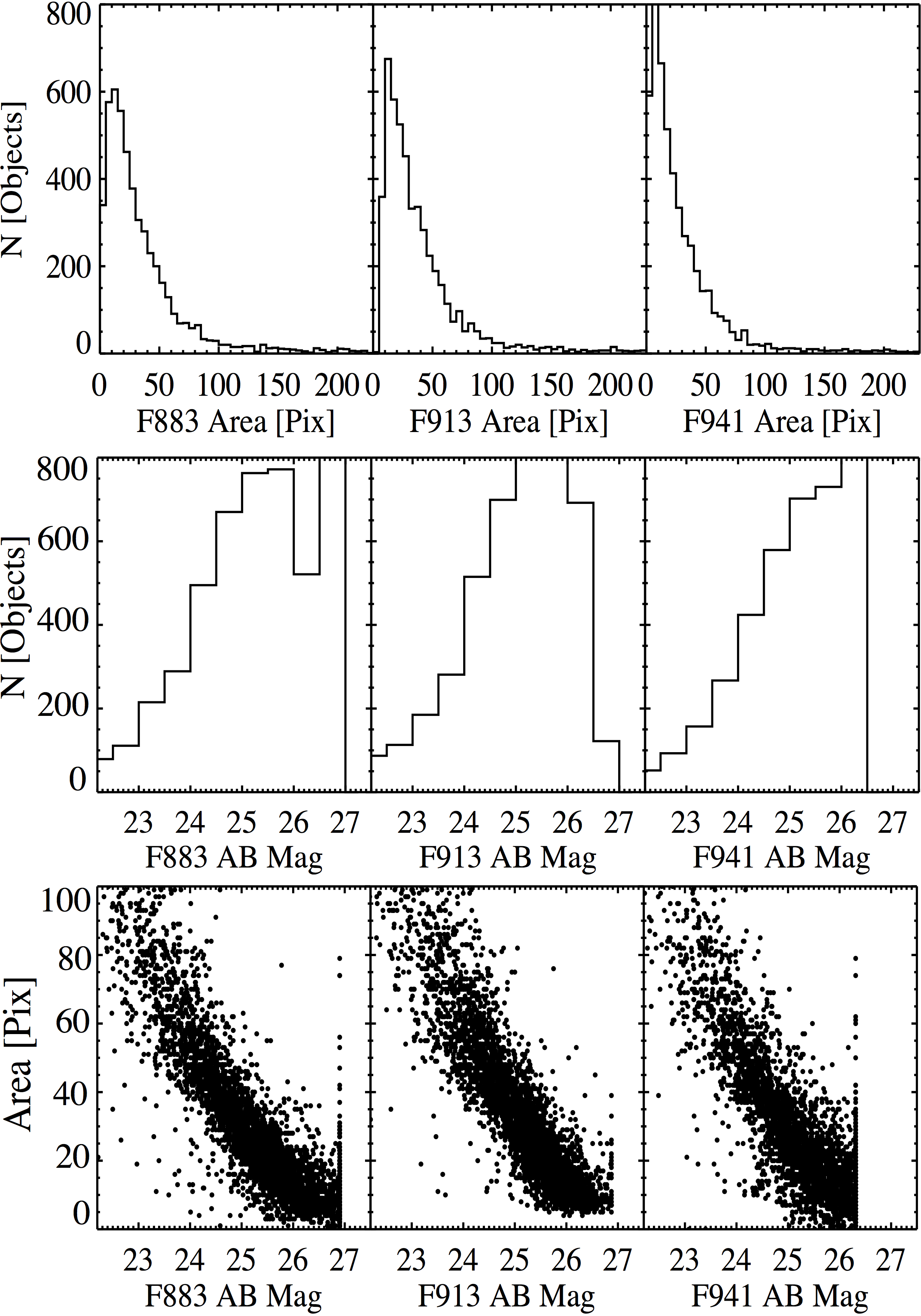}
    \caption{Statistics of objects detected using SExtractor
with the optimised parameters in all 3 bands. 
The left, middle, and right panels represent the statistic of SExtractor isophotal area and auto-magnitudes of the F883, F913, and F941 images, respectively .
The abscissa in the top panel represents the isophotal area of the sources.
The abscissa in the middle and bottom panels represents the auto-magnitudes, assigned from SExtractor then reassigned all non-detections to be 2$\sigma$ magnitude limits.}
    \label{fig:3} 
\end{figure}

		  In order to extract LAE candidates using SExtractor in its dual-image mode, we use the F913 band as the reference image, which is the band covering the redshifted position of Ly$\alpha$ emission at z = 6.5. The positions of objects detected in the F913 band are applied to measure their fluxes and other parameters in the two adjacent bands (i.e., F883 and F941). This procedure is the most robust for targeting LAEs at $z=6.5$. The aperture diameter used for measuring fluxes and aperture magnitude is 2 arcsec. 
		  
		We define the detection threshold in each band to be at 2$\sigma$ magnitudes (i.e., the magnitude that yields signal-to-noise of 2). Objects detected at flux levels lower than 2$\sigma$ limit are assigned 2$\sigma$ magnitudes of the corresponding bands. We adopt SExtractor's auto-magnitudes to be the total AB magnitudes of the objects. However, when calculating colours, we use SExtractor's aperture magnitudes for better assessment of pixel-to-pixel colours. The statistics of the objects detected by SExtractor in dual-image mode using the F913 image as the reference are shown in Figure~\ref{fig:3}. From left to right, the upper, middle, and lower panels of the figure show the area histograms, magnitude histograms, and area-magnitude diagrams for the F883, F913, and F941 bands, respectively. 
		
		Many studies have shown that LAEs are compact objects with half-light radii in the order of a few kpc (e.g.,~\citet{venemans2005},~\citet{Finkelstein2011}, and~\citet{gronwall2011}). Thus, we are paying attention toward the small area and faint tail of the area-magnitude diagram in Figure~\ref{fig:3}. However, we cannot ignore the other region of the area-magnitude diagram, since high redshift LAEs may come in different sizes and luminosities as shown in some previous studies (e.g.~\citet{ota2008, ouchi2008, himiko, ouchi2010}).

\subsection{Candidates Selection}

		The first step in the candidates selection process is to apply colour and magnitude criteria from the expected colour of LAEs at z=6.5. LAEs are categorised as exhibiting a strong and well-distinguished Ly$\alpha$ emission line at 1216 $\AA$ in rest-frame. The region blueward of the Ly$\alpha$ line is obscured by neutral hydrogen gas from the lower-z Intergalactic Medium. Redward of the emission line one could find strong UV continuum in case of the LAEs' higher Star Formation Rate (SFR) cousins, LBGs. Thus, for the purpose of selecting LAE candidates, we look for a clear detection in the F913 band, where the redshifted Ly$\alpha$ emission line should be, and a marginal to non-detection in the F883 and F941 bands. However, the depth of the F941 band (aiming for UV-continuum at z=6.5) is about 0.5 magnitude shallower than the other 2 bands, thus diminishing our ability to distinguish between LAEs and LBGs, the latter exhibit stronger UV-continuum than the former. Due to this reason, we group all the possible LAEs and LBGs in the same list. Then, the detection criteria are revised to optimise the detection of both the LAE candidates and the undistinguishable LBG candidates by allowing some detection in the F941 band up to the same signal-to-noise level as in the F913 band. However, these colour criteria may be fairly shallow and could allow some interlopers with pure red-rising power-law SED, such as high-z quasars (e.g.,~\citet{berk2001}, and~\citet{fan2004}), to be detected as well. Therefore, we conduct a simulation to find the colour cuts that could minimise the contamination from such interlopers. The slope of the power-law SED has to be such that it is steep enough to appear red on F883-F913 colour and without significant detection on F883 band; but, at the same time, shallow enough to appear blue on F913-F941 colour. It turns out that adjusting the colour criteria to be as shown Equation~\ref{eq:color1},~\ref{eq:color2}, and~\ref{eq:color3}, would prevent any object exhibiting power-law SED with slope $\alpha \geq 0.0$ to be detected.

\begin{equation}
 F913(mag-auto) \leq F913(3\sigma \:mag-auto) 
	\label{eq:color1}
\end{equation}

\begin{equation}
    F883 - F913 \geq 0.60 
	\label{eq:color2}
\end{equation}

\begin{equation}
    F913 - F941 \leq 0.40
	\label{eq:color3}
\end{equation}

\begin{figure}
	\includegraphics[width=\columnwidth]{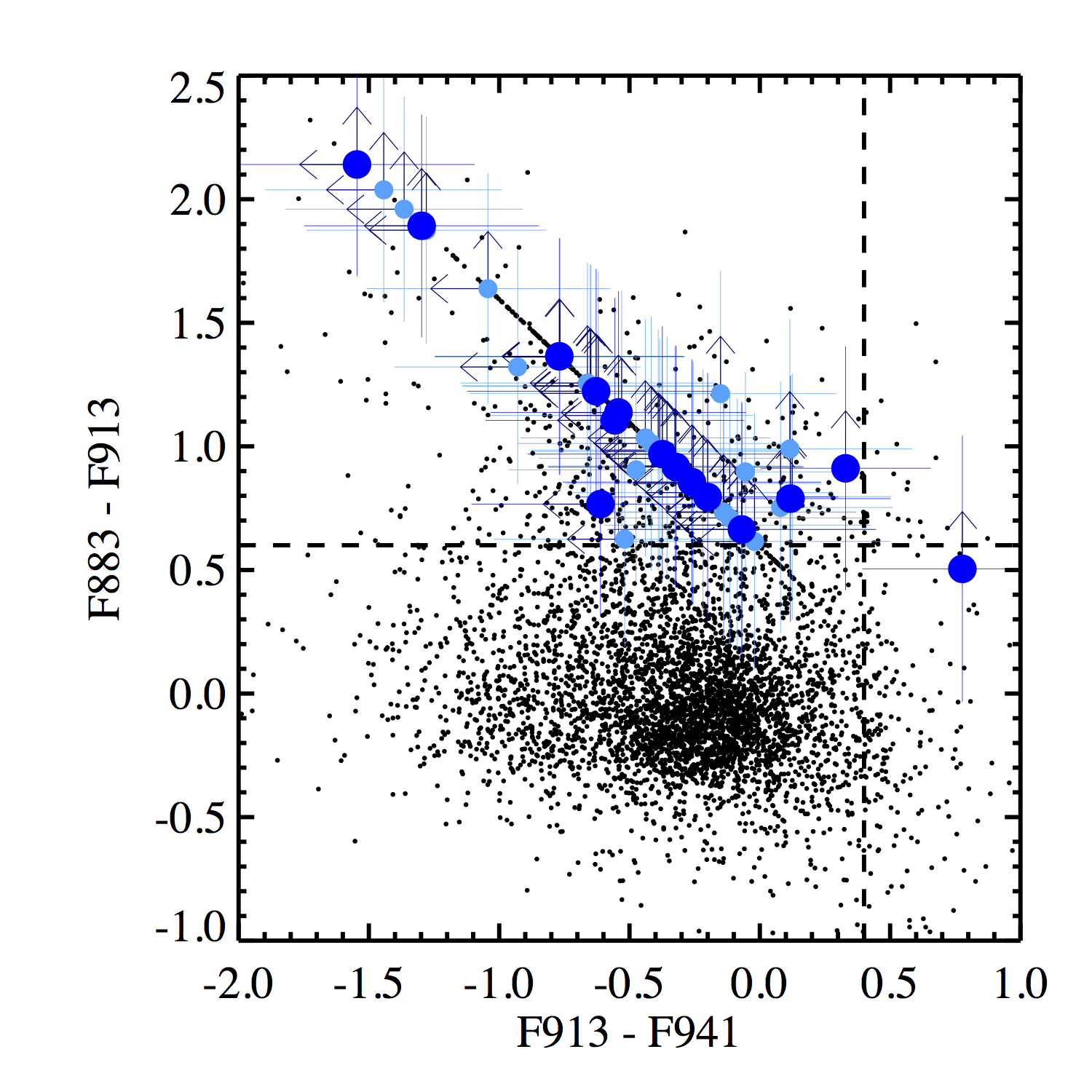}
    \caption{Colour-colour diagram. The objects' colours are calculated using the aperture magnitudes, in order to get the most accurate pixel-to-pixel colours of the objects. We use the aperture size of 2 arcsec to measure the aperture magnitudes of the objects. The vector and abscissa of the diagram represent F883 - F913 colour and F913 - F941 colour. The blue and sky-blue circles represent class-I and -II LAE candidates (classification of LAE candidates is discussed toward the end of section 3.2), respectively.}
    \label{fig:4}
\end{figure}

	The diagram in Figure~\ref{fig:4} shows that the colour criteria are robust and effective for selecting LAEs at z=6.5. In colour-colour space, the diagram shows that the positions of LAE candidates are clearly separated from the majority of the sources, which have both F883-F913 and F913-F941 values clumped around zero. The only LAE candidate that does not meet the criteria, but included in the catalogue is LAE-C-1-02. It is shown in Figure~\ref{fig:4} at F883-F913=0.5 and F913-F941=0.8. We include this candidate in the catalogue, because it is a spectroscopically confirmed LAE at z=6.5, NB921-N-77765~\citep{ouchi2010}. The reason this particular candidate fails the colour criteria is due to the effects of spatially dependent contamination and completeness levels, which we justify in section 3.3.
		
\begin{figure}
	\includegraphics[width=\columnwidth]{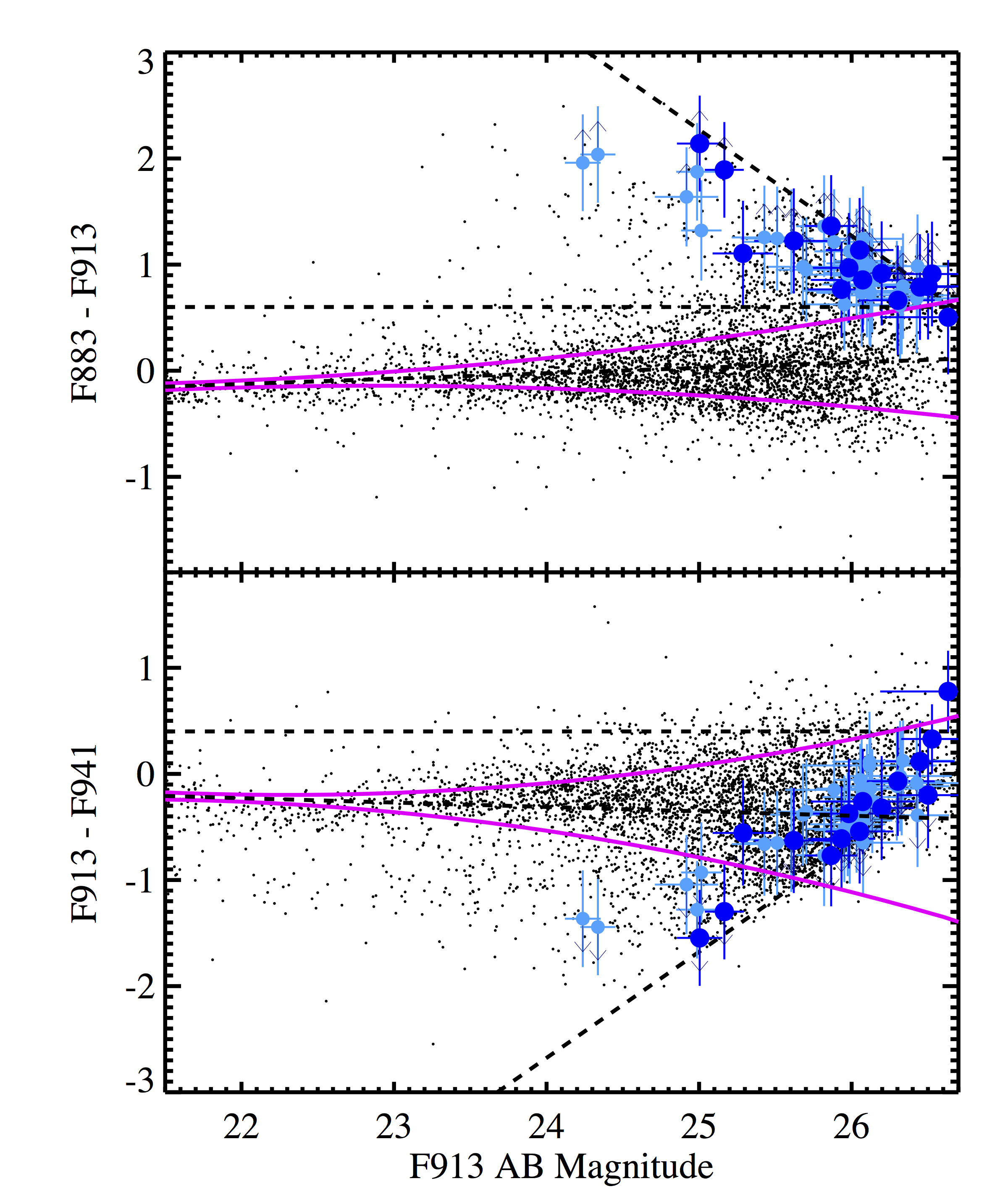}
    \caption{Colour-Magnitude diagrams, adopting aperture magnitudes. The magenta solid lines indicate the 2$\sigma$ boundaries of the uncertainty in F883 - F913 (top panel), and F913 - F941 (bottom panel) colour measurements as functions of AB magnitudes. The middle black dashed lines trace the median colour of the objects in each magnitude bin. The horizontal black dashed lines indicate the colour cut criteria (F883-F913 $\geq$ 0.60 and F913-F941 $\leq$ 0.40). The diagonal black dashed lines indicate the colour boundaries arising from the F883(2$\sigma$) - F913 and F913 - F941(2$\sigma$) colours. Note that the candidates qualify as clearly detected at the $3\sigma$ level within a 2-arcsec aperture. The symbols are as described in Figure~\ref{fig:4}.
    }
    \label{fig:5}
\end{figure}

	 The colour-magnitude diagrams are illustrated in Figure~\ref{fig:5}. The F883-F913 colours of the LAE candidates are beyond the 2$\sigma$ uncertainty in colour from the median colour distribution of all objects. The contours are drawn from the $2\sigma$ uncertainty of the objects' colour measurements, based on standard errors propagation of the aperture magnitude in each band, away from the median colour in each magnitude bin (bin size of 0.1 mag). However, the flux sensitivity of the F941 image is about 0.5 magnitude shallower than the F913 image. The lack of depth in the F941 band only prevents us from distinguishing between LAE and LBG candidates, but not from selecting the star-forming galaxies at z=6.5 from the low-z interlopers. The diagonal line formed by the majority of the LAE candidates in colour-colour diagram in Figure~\ref{fig:4} corresponds to the objects with non-detection in F883 and F941.

\begin{figure}
	\includegraphics[width=1.05\columnwidth]{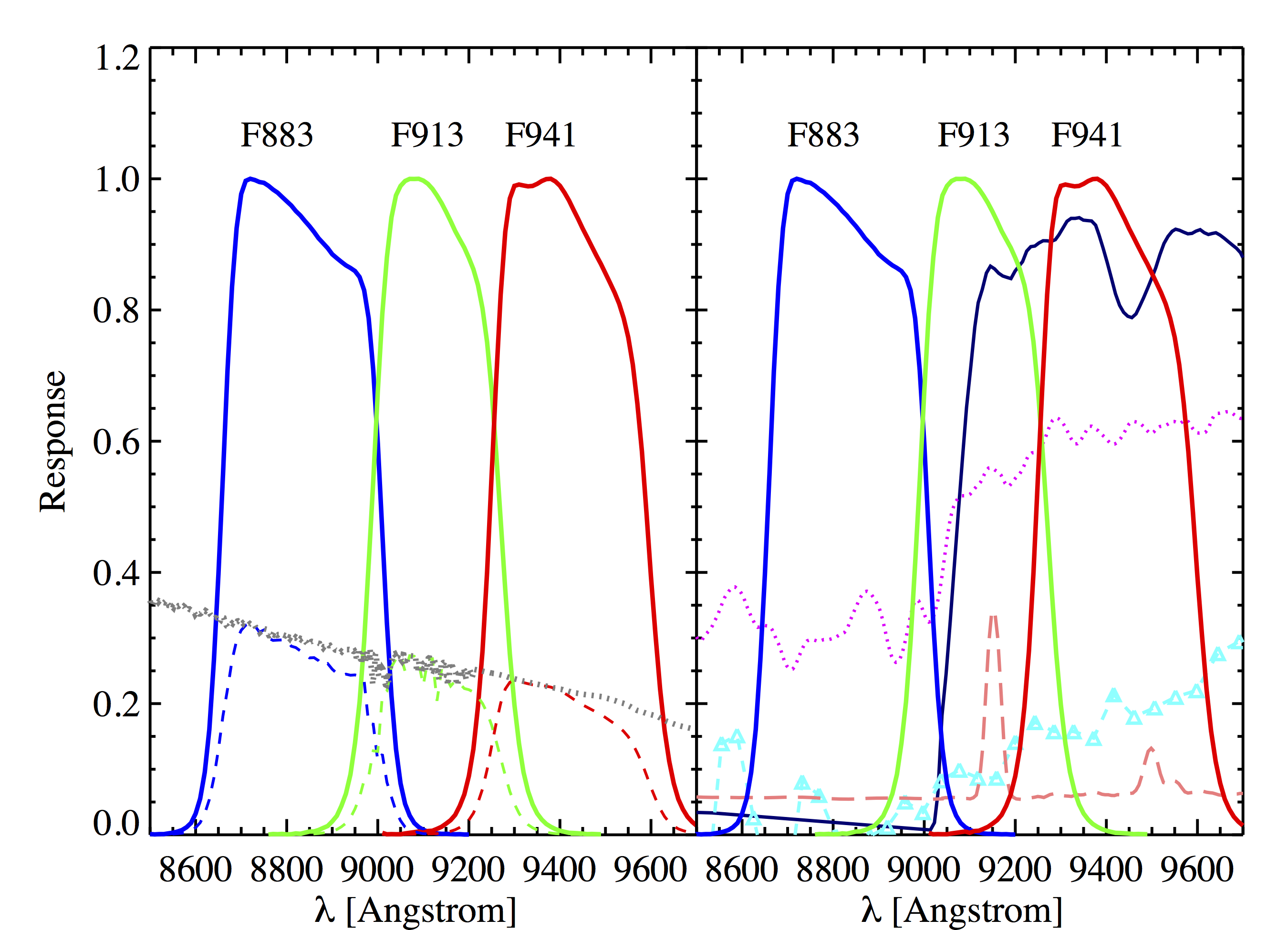}
    \caption{Filter response functions and model SEDs for various type of possible candidates and interlopers.
    \textbf{Left panel:} Blue, green, and red solid lines represent F883, F913, and F941 normalised filter response functions,
respectively. The grey dotted line is the overall throughput of the
system (taking into account total optical throughput of GTC/OSIRIS,
CCD quantum efficiency, and atmospheric transmission). 
\textbf{Right panel:}
Again, the normalised filter response functions are shown in the dashed
lines, overlaid by the various model SEDs of potential sources. 
The navy blue solid line represents the SED of a dropout galaxy~\citep{papovich2001}, and modified transmitted flux density blueward of $Ly\alpha$ to be of what expected of at z=6.5 (extrapolated from~\citet{madau1995});
the magenta dotted line represents the SED of a Balmer-break galaxy~\citep{coleman1980}, a nucleus of an old elliptical galaxy, redshifted to z$\sim$0.4;
the cyan triangle-dashed line represents the SED of a red Galactic dwarf star with spectral type T0~\citep{gunn1983, knapp2004};
and the orchid dashed line represents a dusty starburst galaxy~\citep{cimatti2002}, redshifted to z$\sim$1.4.}
    \label{fig:6}
\end{figure}

\begin{figure}
	\includegraphics[width=1.025\columnwidth]{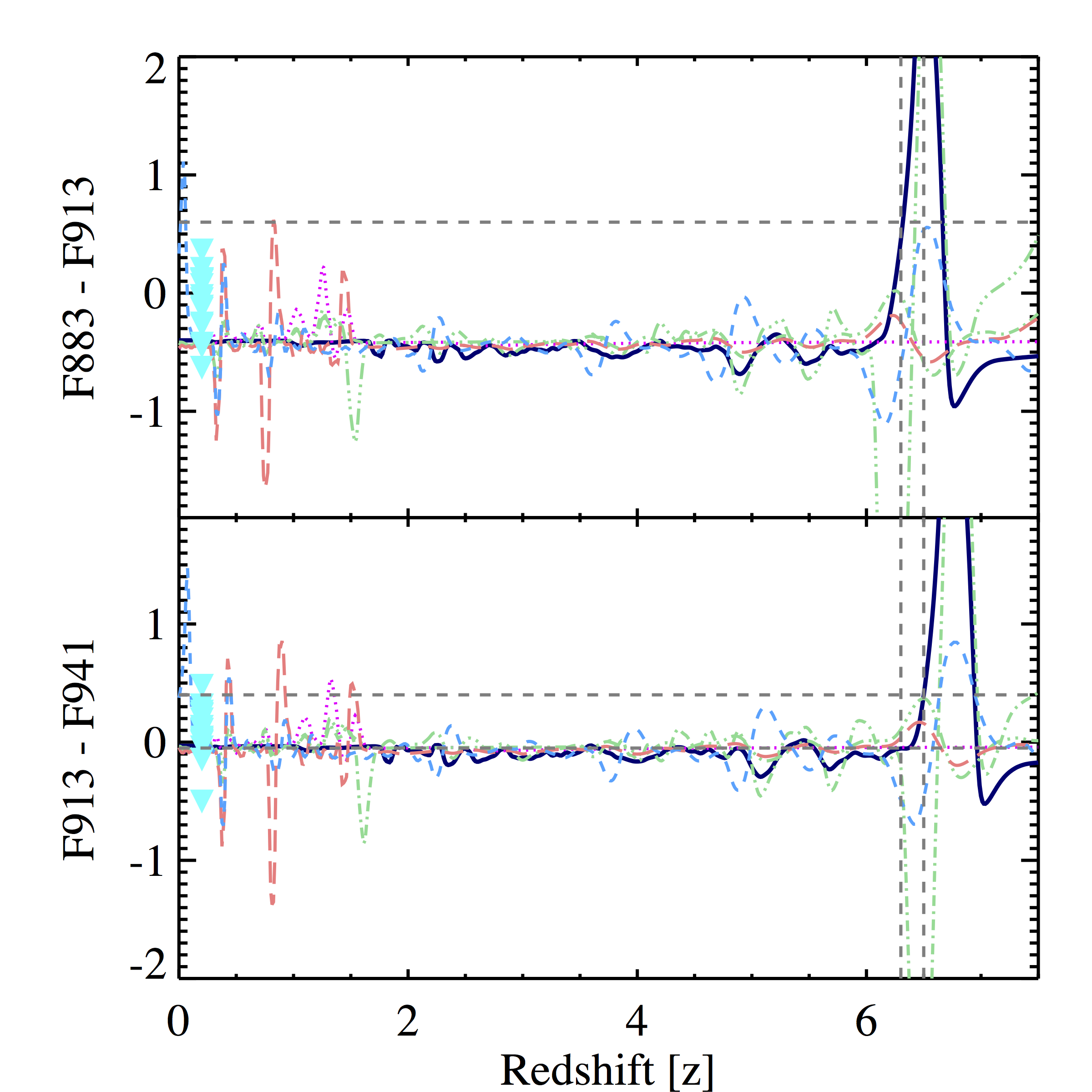}
    \caption{Expected colours as a function of redshift for various type of possible candidates and interlopers.
The navy blue solid, magenta dotted, and orchid long-dashed lines are the expected colours of a dropout galaxy, a Balmer-break galaxy, and a dusty starburst galaxy as functions of redshift.
The green dash-dotted lines are the expected colours of late-type spiral galaxies (e.g.~\citet{terlevich2002}) as a function of redshift. 
The blue dashed lines are the expected colours of high-z quasars derived from the composite SED of SDSS quasars~\citep{berk2001}
The cyan triangles are the expected colours of Galactic L-T dwarfs at z=0.
Notice that we have shifted the abscissa values of the cyan triangles to z=0.2 to avoid plotting the symbols over the Y-axis line. 
From this illustration, the expected colours for dropout galaxies
at z$\sim$6.5 should be as follows: F883-F913$\geq$0.60
and F913-F941$\leq$0.40 for LAEs (with strong Lyman alpha emission and
weak UV continuum).}
    \label{fig:7}
\end{figure}

	To gain a better understanding on how we can truly optimise the colour criteria to include most LAEs with minimum number of low-z interlopers, we conduct a simulation for the expected colour responses from the high-z dropout galaxies and possible interlopers. The first step of the simulation is computing the total GTC/OSIRIS system throughput, taking into account the atmospheric transmission and the filter responses of all 3 bands. We obtain the system magnitude for each band from the products of these filters' total throughputs and the normalised model SEDs of a typical high-z dropout galaxy and possible low-z interlopers at redshift ranges from z=0-7. The simulated SEDs are shown in Figure~\ref{fig:6}. The colour response for the model SEDs of the dropout galaxy and low-z interlopers, including Galactic L-T dwarfs, dusty starburst galaxy, late-type spiral galaxies, and Balmer-break galaxy are calculated from the difference between the calculated system magnitudes. Figure~\ref{fig:7} shows the expected F883-F913 and F913-F941 colours for the high-z dropout galaxy and low-z interlopers as function of redshift. The SED of the model dropout galaxy is set to have almost constant maximum flux density beyond the rest-frame wavelength of 1216 $\AA$, and only 2\% or less of the maximum flux density level for the rest-frame wavelength less than 1216 $\AA$. Doing so, we mimic the SED of LBGs with the strongest possible UV continuum. Thus, the colour response for the high-z dropout galaxy shown in Figure~\ref{fig:7} resemble the most extreme case of LBGs located at z=0-7. 
	
	The expected colours of LBGs are used as boundaries for the expected colours of LAEs. One of the main differences in the SEDs between an LAE and LBG is that the latter has strong rest-frame UV-continuum, while the former does not. Therefore, due to the relative flux densities of the Ly$\alpha$ emission and optically thick Ly$\alpha$ forest, the F883-F913 colour of any z=6.5 LAE should be greater than that of the model of a dropout galaxy. On the other hand, due to the relative flux densities of the Ly$\alpha$ emission and UV-continuum, the F913-F941 colour of any z=6.5 LAE should be lower than that of the dropout galaxies. Thus, for the redshift range approximately between 6.4 to 6.6, the optimised colours that yield the lowest contamination from the low-z interlopers (e.g., H$\alpha$ emitters, [OII-3727] emitters, Balmer break galaxies, and L-T dwarfs) are F883-F913$\geq$0.6 and F913-F941$\leq$0.4. Thus, the simulation results are in good agreement with our pre-determined colour criteria for the LAE candidates at z=6.5. These colour criteria also aid the detection of fainter LAEs at z$\sim$6.5, in comparison to the work by~\citet{ouchi2010}. However, with the concern that the colour criteria are shallower than the usual LAE surveys, we have conducted a simulation to see whether pure power-law SED with arbitrary slope ($0\leq \alpha \leq 10$) would pass such criteria. We found that only a specific range of power-law slope would pass the colour criteria (i.e., $1.7\leq \alpha \leq 2.0$). Furthermore, the simulation of the colour response also shows that the LAE candidates with $F883-F913 \geq 0.6$ and $F913-F941\leq 0.4$, would not be contaminated by power-law SED objects, regardless of the slope.

\begin{figure*}
	\includegraphics[width=1.75\columnwidth]{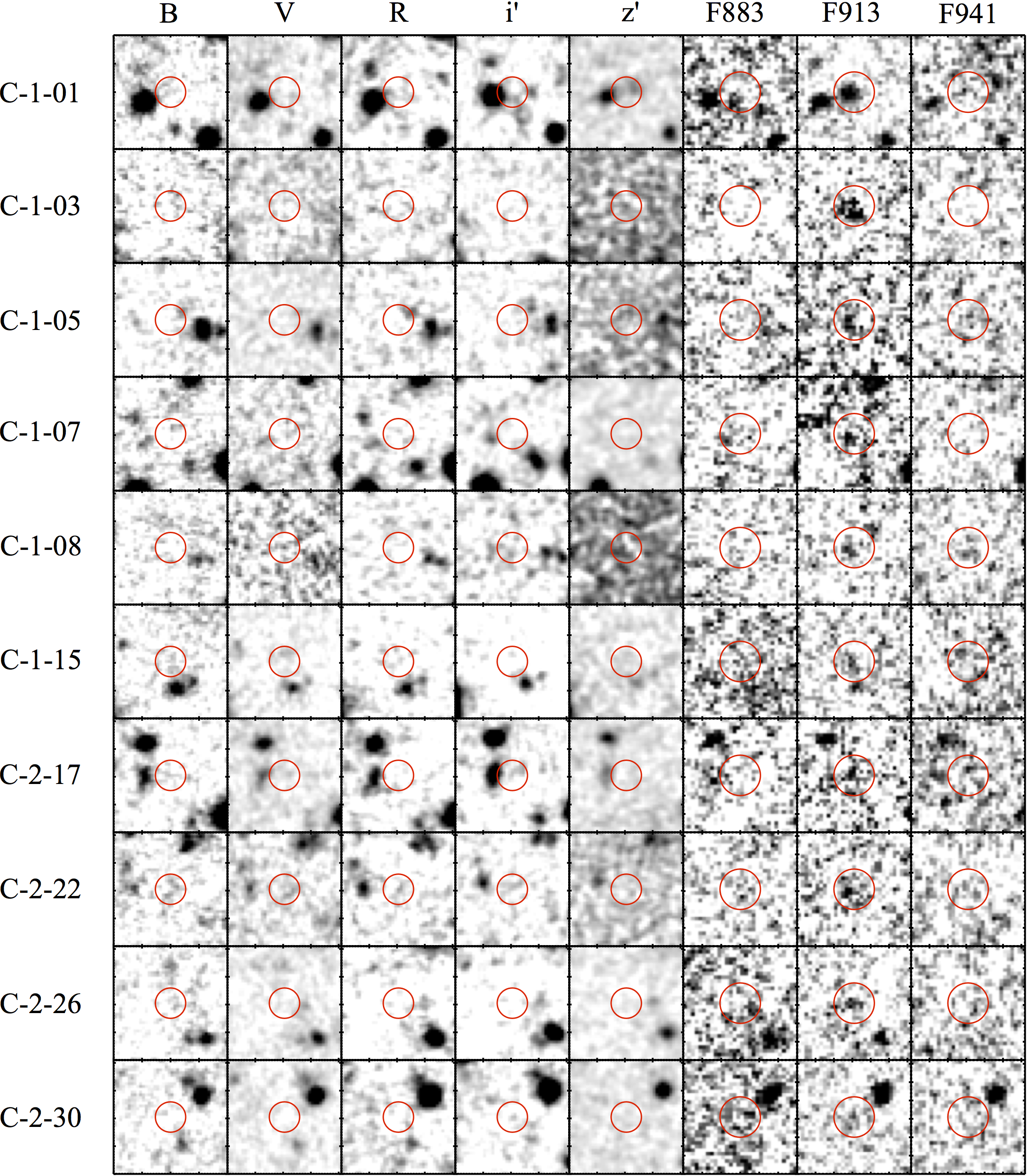}
    \caption{Sample stamp images of LAE candidates both in OSIRIS/SHARDS' medium band filters and SXDS' B, V, R, i', and z' filters. Note that LAE-C1-01 (shown in line 1) is the same as the spectroscopically confirmed massive LAE in SXDS-N from~\citet{ouchi2010}. C-1 denotes class-I (lines 1-6) and C-2 denotes class-II (lines 7-10).}
    \label{fig:8}
\end{figure*}

	All the objects that pass the colour criteria are cross-checked with the SXDS photometric catalogue~\citep{furusawa2008} to check for detections in the B,V, R, i' , or z' bands. Only detections in the z'-band are tolerated, because a strong Ly$\alpha$ emission line at z=6.5 ($\lambda$ = 9120 $\AA$) could be marginally detected in the z' band, covering $\sim$8500-9500 $\AA$. First, we search for non detection in B,V, R, and i' bands using SXDS catalogs and $3\sigma$ limiting magnitudes. Then, we conduct visual inspection through all the 6$\times$6 arcsec$^2$ stamp images of the preliminary LAE candidates and reject those with any sign of marginal detections in the B, V, R, or i' bands. Therefore, the final set of LAE candidates only shows clear detections in F913, marginal to non-detections in z' band, and non-detection in any other bands, as illustrated in Figure~\ref{fig:8}, where we show stamp images of 10 LAE candidates. With the thorough examination, the final set of LAE candidates that pass all of the criteria are only amount to 10\% of the initial selection. 
	
	We categorise candidates into different classes based on their F913 flux profiles. Class-I LAE candidates are those that exhibit flux profiles resembling compact galaxies in F913 band, have peak flux at the centre, and compact almost circular shape. Class-II LAE candidates are those that exhibit flux profiles resembling compact galaxies, but have noise contamination and/or the position of their peak flux is skewed from the centre. Class-III LAE candidates exhibit questionable flux profiles associated with noise fringes, or resemble the appearances of cosmic rays. However, class-III LAE candidates are prone to be spurious detections or residual cosmic rays, rather than actual LAEs at z=6.5. We only catalogue and focus our analysis on class-I and -II LAE candidates.

	  The object IDs, coordinates, along with physical parameters of 47 LAE candidates (2 spectroscopically confirmed LAEs + our 45 LAE candidates) are listed in Table~\ref{tab:2}. The catalogue contains 15 class-I LAE candidates, named LAE-C-1-01 through -15. The total of 32 class-II LAE candidates are also included in the catalogue, namely LAE-C-2-16 through -47. Within class-I LAE candidates, LAE-C-1-01 and LAE-C-1-02 are the 2 spectroscopically confirmed LAE at z$\sim$6.5, namely NB921-N-79144 and NB921-N-77765, respectively~\citep{ouchi2010}.
	  
\begin{figure*}
	\includegraphics[width=1.75\columnwidth]{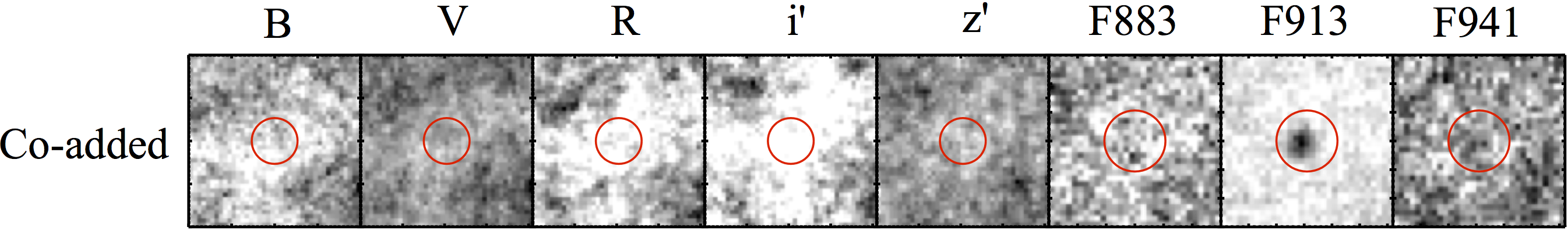}
    \caption{Stack images of 45 class-I and -II LAE candidates, excluding the spectroscopically confirmed massive LAEs from~\citet{ouchi2010}. One can clearly see that there is no significant detection in any other bands except in SHARDS F913, which is the position of the redshifted Ly$\alpha$ emission. These co-added images truly show the signature of an average LAE at z=6.5.}
    \label{fig:9}
\end{figure*}

\begin{figure}
	\includegraphics[width=\columnwidth]{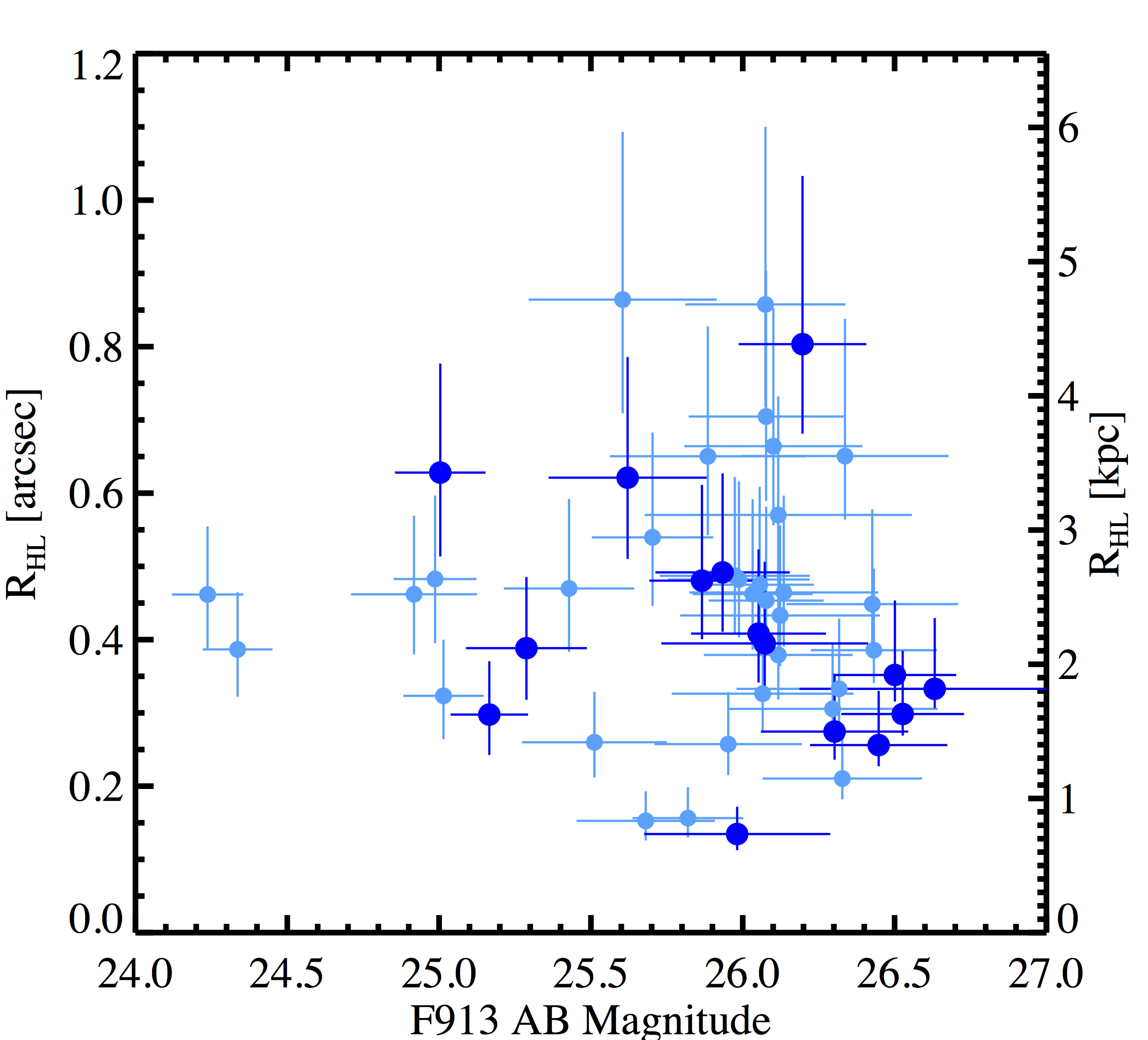}
    \caption{Half-light radii, $R_{HL}$, of $Ly\alpha$ emission regions calculated with a simple PSF spread correction.
The median FWHM of unsaturated point-sources (18-22 magnitude stars) in F913 band is 0.80 arcsec. Note that the observations were under seeing limited regime, with seeing $\sim0.7''$ at the wavelength of 900 nm and median airmass of 1.2. The $R_{HL}$ of the candidates are also shown in $kpc$, assuming that all candidates are located at $z=6.5$.}
    \label{fig:10}
\end{figure}

	  	 The co-added stamp images of the 45 LAE candidates (excluding the 2 confirmed LAEs) in B, V, R, i', z', F883, F913, and F941 bands are shown in Figure~\ref{fig:9}. The very deep broadband photometry, with $3\sigma$ AB magnitudes in B, V, R, i', and z' bands of 28.6, 27.8, 27.7, 27.7, and 26.6 mag, is obtained from the data release of the SXDS survey~\citep{furusawa2008}. The stacked images show non-detection everywhere except in the F913 band, which is the position of the redshifted Ly$\alpha$ emission. The other characteristic of LAEs at high redshift is their compactness, with half-light radii ranging from 0.5 to $\sim$4.0 kpc (e.g.~\citet{venemans2005},~\citet{pirzkal2007},~\citet{bond2009},~\citet{guaita2015},~\citet{momose2014},~\citet{Finkelstein2011},~\citet{gronwall2011}, and~\citet{malhotra2012}). In Figure~\ref{fig:10}, the diagram of Point Spread Function (PSF) corrected half-light radii and F913 AB magnitudes of the LAE candidates are presented. The PSF corrected half-light radii of the LAE candidates in arcsec shown in the figure are calculated by $R_{HL}=\sqrt{R_{HL_{obs}}^{2} - FWHM_{unsat}^{2}}$; where $R_{HL_{obs}}$ and $FWHM_{unsat}$ are the observed half-light radius and the measured FWHM of an unsaturated stellar PSF in arcsec, respectively. The majority of LAE candidates have $R_{HL}\:\leq$ 4 kpc, as expected. 
   
\begin{table*}
	\centering
	\caption{List of LAE candidates.}
	\label{tab:2}
	 \begin{tabular}{lcccccccccc}	      	        
		\hline
		\hline
		$\:\:\:\:\:\:$Object  &  $\alpha$(J2000) &  $\delta$(J2000) &  F883         & F913          &         F941                  &   mag$_{Ly\alpha}$   &      L$_{Ly\alpha}$          &  R$_{HL}$    &     $C_{F913}$     &  $S_{F913}$        \\
		            &                             &                            &       (mag) 	  &        (mag)   &         (mag) 	            &             (mag)             &     ($10^{\:43}$ erg s$^{-1}$)        &      (kpc)         &         &         \\
	       $\:\:\:\:\:\:\:\:\:\:$(1)        & 	  (2)		      & 		(3)           &	      (4)       & 	(5)             &       (6)		              & 		(7)	            &           	(8)                                 &	     (9)   & (10)   &   (11)    	    \\
		\hline	       
LAE-C-1-01* &  2:18:27.0300 & -4:35:08.267  &      $>$27.27  &      25.38 &    $>$26.68 &      25.17 &       1.13 $\pm$ 0.14 &       1.62 $_{-0.30}^{+0.40}$&    0.86 &	0.12		 \\
LAE-C-1-02* &  2:18:23.5437 & -4:35:24.144  &     $>$27.27 &      26.77 &      $\:\:\:$25.99 &  26.63 &       0.30 $\pm$ 0.14 &       1.81 $_{-0.14}^{+0.52}$&     0.40  &	0.17		\\   
LAE-C-1-03 &  2:18:07.5897 &  -4:36:46.399 &      $>$27.27 &      25.13 &     $>$26.68 &        25.00  &      1.31 $\pm$ 0.19  &      3.43 $_{-0.62}^{+0.81}$ &     0.92   &	0.09		 \\
LAE-C-1-04 &  2:18:08.0200 &   -4:30:28.145 &      $>$27.27 &      25.91 &      $>$26.68 &      25.87  &      0.60 $\pm$ 0.10 &       2.62 $_{-0.44}^{+0.71}$ &    0.68    &	0.13		\\
LAE-C-1-05 &  2:18:08.2186 &   -4:38:00.294 &      $>$27.27 &      26.05 &      $>$26.68 &      25.62 &       0.74 $\pm$ 0.18 &       3.39 $_{-0.60}^{+0.90}$ &    0.83   &	0.07		\\
LAE-C-1-06 &  2:18:07.1575 &   -4:37:44.814 &      $>$26.83 &      26.07 &      $>$26.68 &      25.93 &       0.56 $\pm$ 0.12 &       2.68 $_{-0.44}^{+0.74}$ &    0.83   &	0.09		\\
LAE-C-1-07 &  2:18:06.4901 &   -4:32:35.247 &      $>$27.23 &      26.12 &      $>$26.68 &      25.29 &       1.01 $\pm$ 0.19 &       2.12 $_{-0.38}^{+0.53}$ &    0.92    &	0.23		\\
LAE-C-1-08 &  2:18:09.1250 &  -4:32:48.995 &      $>$27.27 &      26.13 &      $>$26.68  &     26.05  &      0.50 $\pm$ 0.11&         2.22 $_{-0.36}^{+0.62}$  &   0.57     &	0.06		\\
LAE-C-1-09 &  2:18:07.2344 &  -4:36:38.999 &      $>$27.27 &      26.30 &      $>$26.68  &     25.98  &      0.53 $\pm$ 0.16 &       0.72 $_{-0.12}^{+0.20}$ &     0.81     &	0.33		\\
LAE-C-1-10 &  2:18:22.8826 &  -4:35:10.499 &      $>$27.27 &      26.35 &      $>$26.68  &     26.20  &      0.44 $\pm$ 0.09 &       4.38 $_{-0.66}^{+1.25}$ &     0.56    &  0.06		\\
LAE-C-1-11 &  2:18:22.4661 &  -4:36:54.651 &      $>$27.27 &      26.36 &      $\:\:\:$26.03 &  26.53  &      0.32 $\pm$ 0.06 &       1.62 $_{-0.16}^{+0.47}$ &     0.38   & 	0.09  	 \\
LAE-C-1-12 &  2:18:23.4265 &  -4:30:20.084 &      $>$27.27 &      26.48 &      $>$26.68  &     26.50  &      0.33 $\pm$ 0.07 &       1.92 $_{-0.20}^{+0.55}$ &     0.58   & 	0.08		 \\
LAE-C-1-13 &  2:18:26.8276 &  -4:31:21.075 &      $>$27.27 &      26.48 &      $\:\:\:$26.37 &  26.45  &      0.35 $\pm$ 0.08 &       1.39 $_{-0.16}^{+0.40}$ &      0.41   & 0.09	\\
LAE-C-1-14 &  2:18:22.4066 &  -4:33:21.787 &      $>$27.27 &      26.61 &      $>$26.68  &     26.30  &      0.40 $\pm$ 0.09 &       1.49 $_{-0.21}^{+0.43}$ &      0.50   &	0.30	 \\
LAE-C-1-15 &  2:18:22.5997 &  -4:35:27.820 &      $>$27.27 &      26.42 &      $>$26.68  &     26.07  &      0.49 $\pm$ 0.17 &       2.15 $_{-0.35}^{+0.61}$ &      0.59    & 0.06		\\
LAE-C-2-16 &  2:18:08.6727 &  -4:36:53.597 &      $>$27.27 &      25.23 &      $>$26.68  &     24.34  &      2.42 $\pm$ 0.28 &       2.11 $_{-0.35}^{+0.42}$ &      0.94    & 0.32      		\\
LAE-C-2-17 &  2:18:12.1847 &  -4:38:14.007 &     $>$27.27 &       26.03  &     $>$26.68   &    25.51   &     0.82 $\pm$ 0.19  &      1.41 $_{-0.26}^{+0.38}$ &       0.80   &  0.18               \\
LAE-C-2-18 &  2:18:07.0651 &  -4:38:12.875 &      $>$27.27 &      26.03 &      $>$26.68  &     26.07  &      0.49 $\pm$ 0.12 &       4.68 $_{-0.76}^{+1.32}$ &        0.81  &   0.12              \\
LAE-C-2-19 &  2:18:07.0852 &  -4:37:10.125 &      $\:\:\:$26.78 &  26.16 &      $>$26.68  &     25.95  &      0.55 $\pm$ 0.13 &       1.40 $_{-0.23}^{+0.39}$ &        0.50  &   0.12              \\
LAE-C-2-20 &  2:18:21.8270 &  -4:32:53.379 &      $>$27.27 &      26.24 &      $>$26.68  &     26.08  &      0.49 $\pm$ 0.09 &       2.47 $_{-0.40}^{+0.70}$ &        0.70  &   0.19              \\
LAE-C-2-21 &  2:18:09.2129 &  -4:37:57.352 &      $>$27.27 &      26.26 &      $>$26.68  &     26.03  &      0.50 $\pm$ 0.10 &       2.52 $_{-0.41}^{+0.71}$ &        0.56  &   0.05              \\
LAE-C-2-22 &  2:18:07.0303 &  -4:37:00.984 &      $\:\:\:$27.07 &  25.75 &      $>$26.68  &     25.01  &      1.30 $\pm$ 0.17 &       1.76 $_{-0.32}^{+0.42}$ &       0.87   &   0.08              \\
LAE-C-2-23 &  2:18:07.1685 &  -4:35:38.490 &      $>$27.27 &      26.02 &      $>$26.68  &     25.43  &      0.88 $\pm$ 0.18 &       2.56 $_{-0.47}^{+0.66}$ &       0.81   &   0.21              \\
LAE-C-2-24 &  2:18:07.8085 &  -4:38:10.968 &      $>$27.27 &      26.15 &      $>$26.68  &     25.99  &      0.53 $\pm$ 0.12 &       2.63 $_{-0.43}^{+0.73}$ &       0.73   &   0.09              \\
LAE-C-2-25 &  2:18:21.8096 &  -4:36:59.126 &      $>$27.27 &      26.29 &      $>$26.68  &     26.43  &      0.35 $\pm$ 0.07 &       2.10 $_{-0.24}^{+0.61}$ &       0.34   &   0.10              \\
LAE-C-2-26 &  2:18:22.4203 &  -4:37:12.593 &      $>$27.27 &      26.32 &      $>$26.68  &     25.70  &      0.68 $\pm$ 0.13 &       2.94 $_{-0.51}^{+0.78}$ &       0.76   &   0.06              \\
LAE-C-2-27 &  2:18:26.6885 &  -4:34:15.614 &      $>$27.27 &      26.35 &      $>$26.68  &     26.08  &      0.49 $\pm$ 0.12 &       3.84 $_{-0.63}^{+1.08}$ &       0.72   &   0.05              \\
LAE-C-2-28 &  2:18:23.1015 &  -4:30:55.541 &      $>$27.27 &      26.52 &      $\:\:\:$26.44 &  26.12  &      0.47 $\pm$ 0.22 &       3.11 $_{-0.50}^{+0.88}$ &        0.66  &    0.15             \\
LAE-C-2-29 &  2:18:25.1010 &  -4:31:02.156 &      $>$27.27 &      26.54 &      $>$26.68  &     26.12  &      0.47 $\pm$ 0.15 &       2.36 $_{-0.38}^{+0.67}$ &       0.65   &    0.07             \\
LAE-C-2-30 &  2:18:24.8428 &  -4:37:11.100 &      $>$27.27 &      26.56 &      $>$26.68  &     26.30  &      0.40 $\pm$ 0.14 &       1.66 $_{-0.23}^{+0.48}$ &        0.60  &    0.08             \\
LAE-C-2-31 &  2:18:09.5141 &  -4:38:28.900 &      $>$27.27 &      25.40 &      $>$26.68  &     24.99  &      1.33 $\pm$ 0.18 &       2.63 $_{-0.48}^{+0.62}$ &        0.69  &    0.07              \\
LAE-C-2-32 &  2:18:07.8607 &  -4:30:29.718 &      $>$27.27 &      25.91 &      $>$26.68  &     25.82  &      0.62 $\pm$ 0.11 &       0.84 $_{-0.14}^{+0.23}$ &        0.72  &    0.07              \\
LAE-C-2-33 &  2:18:06.5917 &  -4:37:59.999 &      $\:\:\:$26.88 &  25.97 &      $\:\:\:$26.45 &  25.97  &      0.54 $\pm$ 0.13 &       2.66 $_{-0.44}^{+0.73}$ &        0.85  &    0.06              \\
LAE-C-2-34 &  2:18:06.9561 &  -4:38:22.085 &      $>$27.27 &      26.28 &      $\:\:\:$26.17 &  26.12  &      0.47 $\pm$ 0.12 &       2.07 $_{-0.33}^{+0.59}$ &        0.78  &    0.05             \\
LAE-C-2-35  & 2:18:06.7520 &  -4:32:22.535 &      $>$27.27 &      25.31 &      $>$26.68  &     24.24  &      2.66 $\pm$ 0.32 &       2.52 $_{-0.42}^{+0.51}$ &        0.95  &    0.10              \\
LAE-C-2-36  & 2:18:06.9735 &  -4:30:34.311 &      $>$27.27 &      26.06 &      $>$26.68  &     25.60  &      0.76 $\pm$ 0.23 &       4.72 $_{-0.84}^{+1.24}$ &        0.69  &    0.07              \\
LAE-C-2-37  & 2:18:06.8142 &  -4:38:03.216 &      $>$27.27 &      26.06 &      $\:\:\:$26.21 &  25.88  &      0.58 $\pm$ 0.19 &       3.55 $_{-0.59}^{+0.96}$ &        0.88  &    0.04              \\
LAE-C-2-38  & 2:18:26.3607 &  -4:34:14.271 &      $\:\:\:$27.11 &  26.22 &      $\:\:\:$26.27  & 26.06 &       0.49 $\pm$ 0.08 &       2.59 $_{-0.42}^{+0.73}$ &        0.71  &    0.06              \\
LAE-C-2-39  & 2:18:32.1844 &  -4:33:42.234 &      $>$27.27 &      26.42 &      $>$26.68  &     26.14 &       0.46 $\pm$ 0.14 &       2.53 $_{-0.40}^{+0.72}$ &        0.86  &    0.18             \\
LAE-C-2-40  & 2:18:28.9901 &  -4:30:45.280 &      $>$27.27 &      26.48 &      $\:\:\:$26.35 &  26.34 &       0.38 $\pm$ 0.13 &       3.55 $_{-0.47}^{+1.02}$ &         0.64 &    0.24              \\
LAE-C-2-41  & 2:18:25.0030 &  -4:31:12.727 &      $>$27.27 &      26.59 &      $>$26.68  &     26.43 &       0.35 $\pm$ 0.10 &       2.44 $_{-0.29}^{+0.70}$ &        0.38  &    0.04              \\
LAE-C-2-42  & 2:18:06.7337 &  -4:30:20.048 &      $>$27.27 &      25.63 &      $>$26.68  &     24.92 &       1.42 $\pm$ 0.29 &       2.52 $_{-0.44}^{+0.58}$ &        0.99   &   0.09              \\
LAE-C-2-43  & 2:18:24.1854 &  -4:35:40.977 &      $>$27.27 &      26.46 &      $>$26.68  &     26.10 &       0.48 $\pm$ 0.14 &       3.62 $_{-0.59}^{+1.03}$ &        0.79  &    0.31              \\
LAE-C-2-44  & 2:18:30.0796 &  -4:33:59.011 &      $>$27.27 &      26.61 &      $>$26.68  &     26.33 &       0.39 $\pm$ 0.09 &       1.14 $_{-0.15}^{+0.33}$ &        0.49  &     0.05             \\
LAE-C-2-45  & 2:18:23.0538 &  -4:32:50.351 &      $>$27.27 &      26.54 &      $>$26.68  &     26.07 &       0.49 $\pm$ 0.15 &       1.78 $_{-0.29}^{+0.50}$ &        0.65  &    0.07            \\
LAE-C-2-46  & 2:18:29.0698 &  -4:36:41.021 &      $>$27.27 &      26.66 &      $>$26.68  &     26.32 &       0.39 $\pm$ 0.13 &       1.81 $_{-0.25}^{+0.52}$ &        0.53  &    0.17             \\
LAE-C-2-47  & 2:18:29.9011 &  -4:30:18.125 &      $>$27.27 &      26.29 &      $>$26.68  &     25.68 &       0.70 $\pm$ 0.16  &      0.82 $_{-0.14}^{+0.22}$ &        0.89  &    0.31             \\
			\hline
	\end{tabular}
	
	\raggedright
	\textbf{Notes.} *The 2 spectroscopically confirmed LAEs, NB921-N-79144 and NB921-N-77765, from~\citet{ouchi2010}. (1) Object ID; (2, 3) R.A. and Declination; (4) F883 aperture magnitude; (5) F913 aperture magnitude; (6) F941 aperture magnitude; (7) F913 AB magnitude; (8) Ly$\alpha$ luminosity in $10^{\:43}$ erg s$^{-1}$; (9) half-light radius in kpc; (10) completeness probability, (11) and spurious probability (i.e., probability of the candidate being spurious detection) corresponding to its F913 magnitude and position on the field.
	
\end{table*}

\subsection{Validity of the Candidates}

\begin{figure}
	\includegraphics[width=1.05\columnwidth]{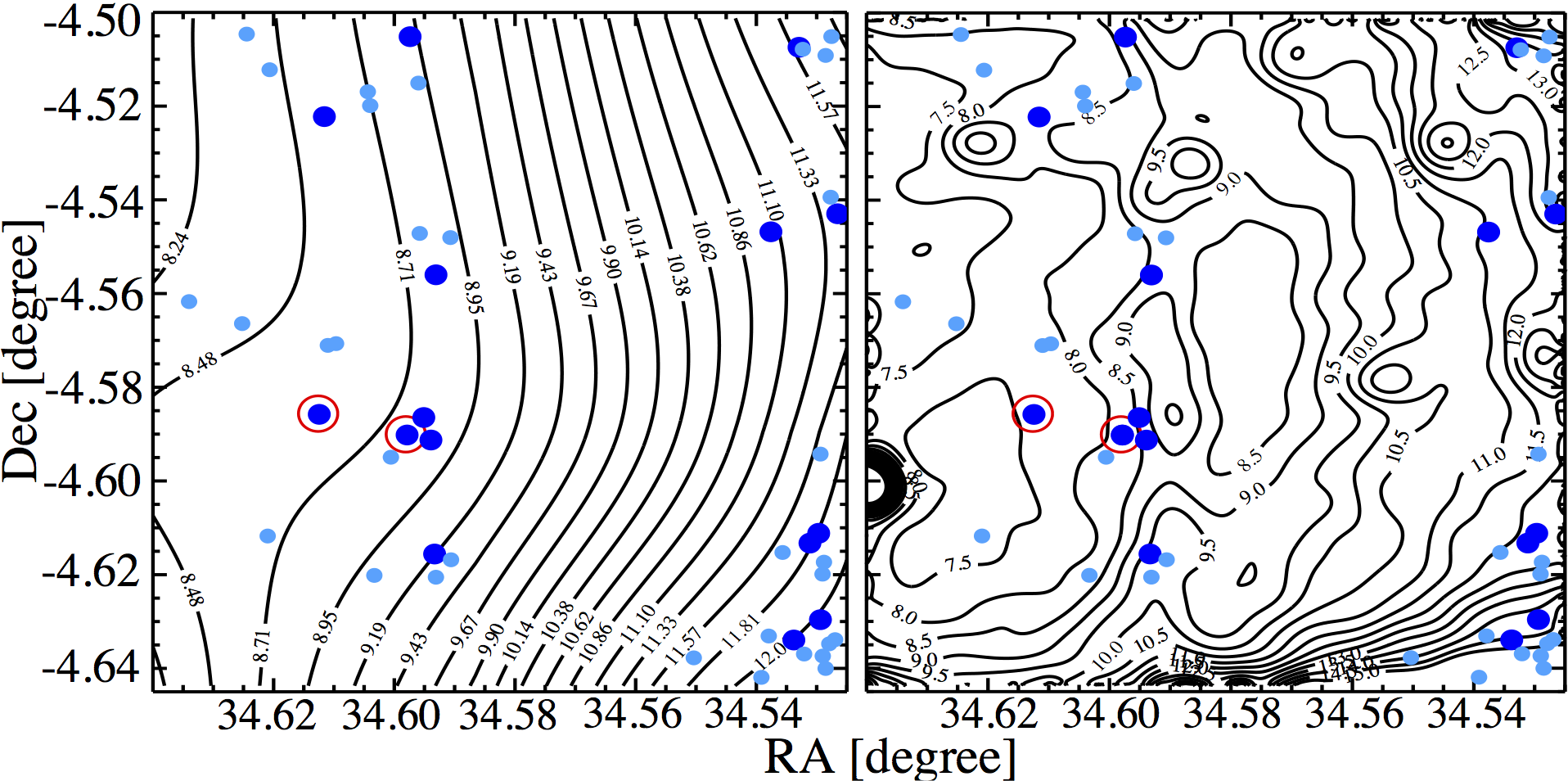}
    \caption{Spatial positions of the final LAE candidates,
overlaid on the contour maps of RMS noise level in ADU. {\bf Left:} The RMS noise contour map made by SExtractor. {\bf Right:} The RMS noise contour map made by our own routine with running $100\times 100 \: pix^2$ sub-region in assessment of background noise. The red circles
indicate the positions of the 2 spectroscopically confirmed LAEs from~\citet{ouchi2010}. 
The plate scale is 0.254''/pixel. The symbols are as described in Figure~\ref{fig:4}.}
    \label{fig:11}
\end{figure}

		The final reduced images in all 3 bands exhibit disparity between noise levels on the left and right sides of the images, corresponding to the left and right OSIRIS CCDs. This is the combined effect of two major factors. The first factor is the wavelength variation across the FOV of the filters used in this survey. From~\citet{pablo2013}, the variation of the filter's central wavelength from the optical axis for SHARDS filter (including F883, F913, F941) is $\sim2.8\times10^{-5} \AA/pixel^2$. Since the OSIRIS operates off-axis for medium band imaging, the wavelength variation could be up to $\sim 100 \AA$ from left to right edges of the detector (about 2000 pixels across). However, in our case, the FOV span only 1640 pixel in x-direction (R.A.) after trimming. Thus the realistic estimation for the wavelength variation would be $\sim 70 \AA$ across the FOV. The mentioned effect makes the NIR night sky fringes more prominent in particular regions of the FOV than the others. The second factor is the difference in quality between the 2 OSIRIS CCDs. The combined effect of these two factors generates a steeply rising RMS noise level on the right CCD (x$_{pix}\geq$ 800 or RA$\leq$2$^h$18$^m$20.25$^s$). This prevents the selection routines from detecting such faint sources like the LAE candidates at z=6.5. The spatial distribution of the LAE candidates and a contour map of the F913 background RMS noise are shown in Figure~\ref{fig:11}. The two red, open circles indicate the positions of the spectroscopically confirmed LAEs from~\citet{ouchi2008, ouchi2010}. 
	
		Due to the disparity of noise and differential contamination levels between the 2 OSIRIS CCDs, we cannot use the traditional integrated completeness and contamination functions for the whole image. Instead, we conduct an analysis to assess completeness and contamination values of each individual pixel for a specific range of magnitudes for all 3 images. Furthermore, the positions of the LAE candidates also exhibit a peculiar bimodal distribution. 2D simulations for completeness and contamination (spurious detection) levels in all 3 bands are needed to gain a better understanding of these effects. We will discuss the benefits of this analysis in the next section. Here, we explain how to construct such a 2D treatment of completeness and contamination levels.
		
		The completeness and contamination simulations are carried out as discussed in the previous section. However, this time, we assign the values of completeness and contamination levels onto each individual pixel of the images. First, we sort the magnitude range into 6 bins, starting from 24.2 mag to 26.6 mag, with 0.4 mag bin size. In each magnitude bin, we calculate the contamination level from the ratio between the numbers of spurious to real detections that fall within 100-pixel radius from the pixel of interest (reference pixel). Similarly, we calculate the completeness level from the ratio between the numbers of simulated objects detected to total simulated objects injected within 100-pixel radius from the reference pixel. For the completeness level, we repeat this recovering process 200 times for each band and magnitude bin to obtain the pixel-to-pixel median values of the completeness levels.
		
\begin{figure}
	\includegraphics[width=1.01\columnwidth]{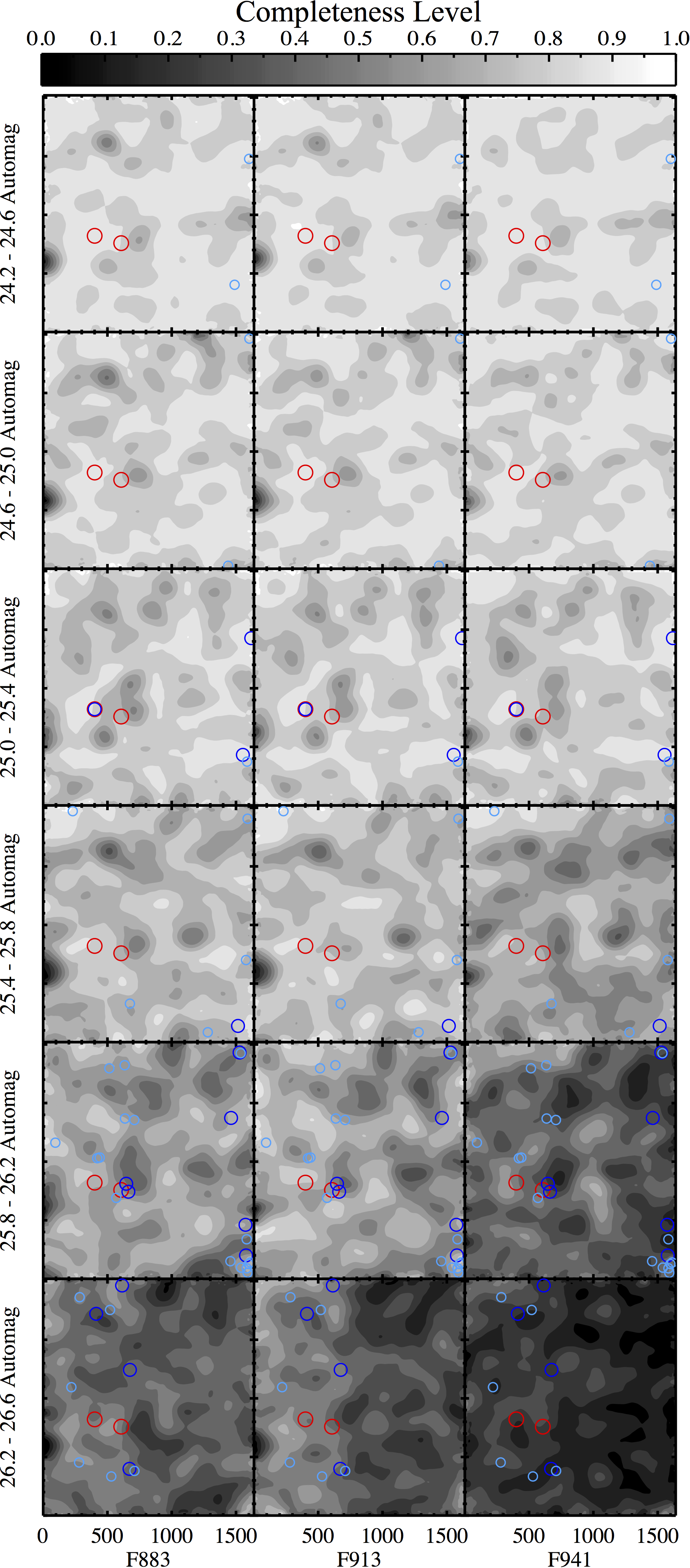}
    \caption{\textbf{From left to right:} Completeness levels in the F883, F913, and F941 bands based on spatial position on the image
and AB magnitude bin. We adopt the colour and symbol codes for all classes of the LAE candidates and detection scenarios. 
The difference between noise levels in the 2 OSIRIS CCDs causes the differential completeness levels across the FOV. 
The concentration of LAE candidates on the left OSIRIS CCD is likely caused by this differential completeness.}
    \label{fig:12}
\end{figure}

\begin{figure}
	\includegraphics[width=1.045\columnwidth]{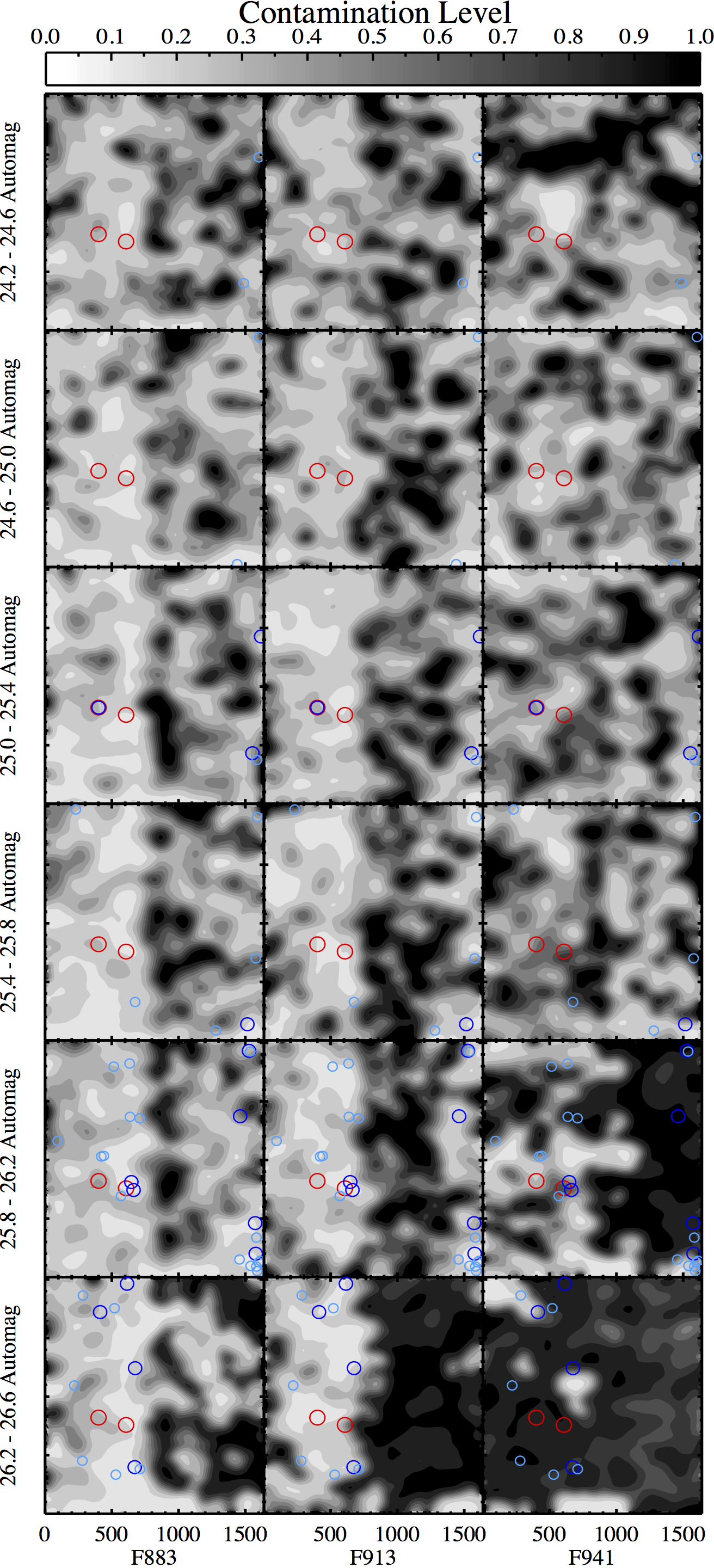}
    \caption{\textbf{From left to right:} Contamination levels in the F883, F913, and F941 bands based on spatial position on the image
and AB magnitude bin. High contamination levels and RMS
noise on the right OSIRIS CCD may prevent the detection of LAEs in that region. 
The positions of LAE candidates are also consistent with the low contamination regions.}
    \label{fig:13}
\end{figure}

		  Figure~\ref{fig:12} and ~\ref{fig:13} are the filled-contour plots for the 2D completeness and contamination levels in each magnitude bin for the three bands. The 25.8-26.2 mag and 26.2-26.6 mag bins clearly show the differences in completeness levels between 2 CCDs of each band. The left OSIRIS CCD clearly exhibits higher completeness levels and much lower contamination levels than those of the right OSIRIS CCD, especially for the last 2 magnitude bins. Overall, the F941 band has the lowest completeness levels. While, the F883 and F913 bands have equally high completeness levels. Nevertheless, the contamination levels of the F883 and F941 bands are higher than those of the F913 band. This proves that the F913 band has advantages in terms of both completeness and contamination levels compared to the other bands. 
		  
		   The 2D completeness and contamination maps can also be used for computing the probability of an LAE candidate being a real detection as a function of magnitude and position on the images. The probability for an LAE candidate being a real detection is derived from the probability of the detection in the F913 image to be real (i.e., not a spurious source), as expressed in Equation~\ref{eq:4}. S$_{F913}$ is the ratio between the numbers of spurious detections to real detections as a function of magnitude and xy-position (i.e., 2D contamination function) on the F913 image. However, if the LAE candidate shows some marginal detection brighter than 2$\sigma$ magnitude in either the F883 or F941 band, Equation~\ref{eq:4} has to be modified by treating those marginal detections as contaminations. Thus, in this scenario, the probability of real detection can be modified by multiplying Equation~\ref{eq:4} with the 2D spurious function from the band and magnitude bin with the marginal detection.

\begin{equation}
 P_{F913}(LAE) = 1- S_{F913}(mag, x, y) 
	\label{eq:4}
\end{equation}
	   		   		   		   
		   Valid LAE candidates should have a probability of real detection $\geq$ 0.5. This extra criterion helps in the validation of the LAE candidates. All the LAE candidates from the final catalogue fall in regions with high completeness and low contamination levels in the F913 band. With this validation and selection of candidates, we have solved the puzzle of the peculiar spatial distribution of the LAE candidates. However, we cannot rule out contamination from possible low-z interlopers. Our probability simulations only serve as a way to select candidates with a good chance of being real detections, but cannot differentiate other galaxies from LAEs at z=6.5. To truly confirm whether the LAE candidates are real, we have to resource to spectroscopic follow-up. Nevertheless, the 2D probability assessment gives us the confidence that we have selected the best candidates in the regions with statistically low contamination and high completeness. This extra step of validation also proves to be useful in target selection for spectroscopic follow-up and statistical decontamination of the integrated properties of the population (i.e., Luminosity Function). 

\begin{figure}
	\includegraphics[width=\columnwidth]{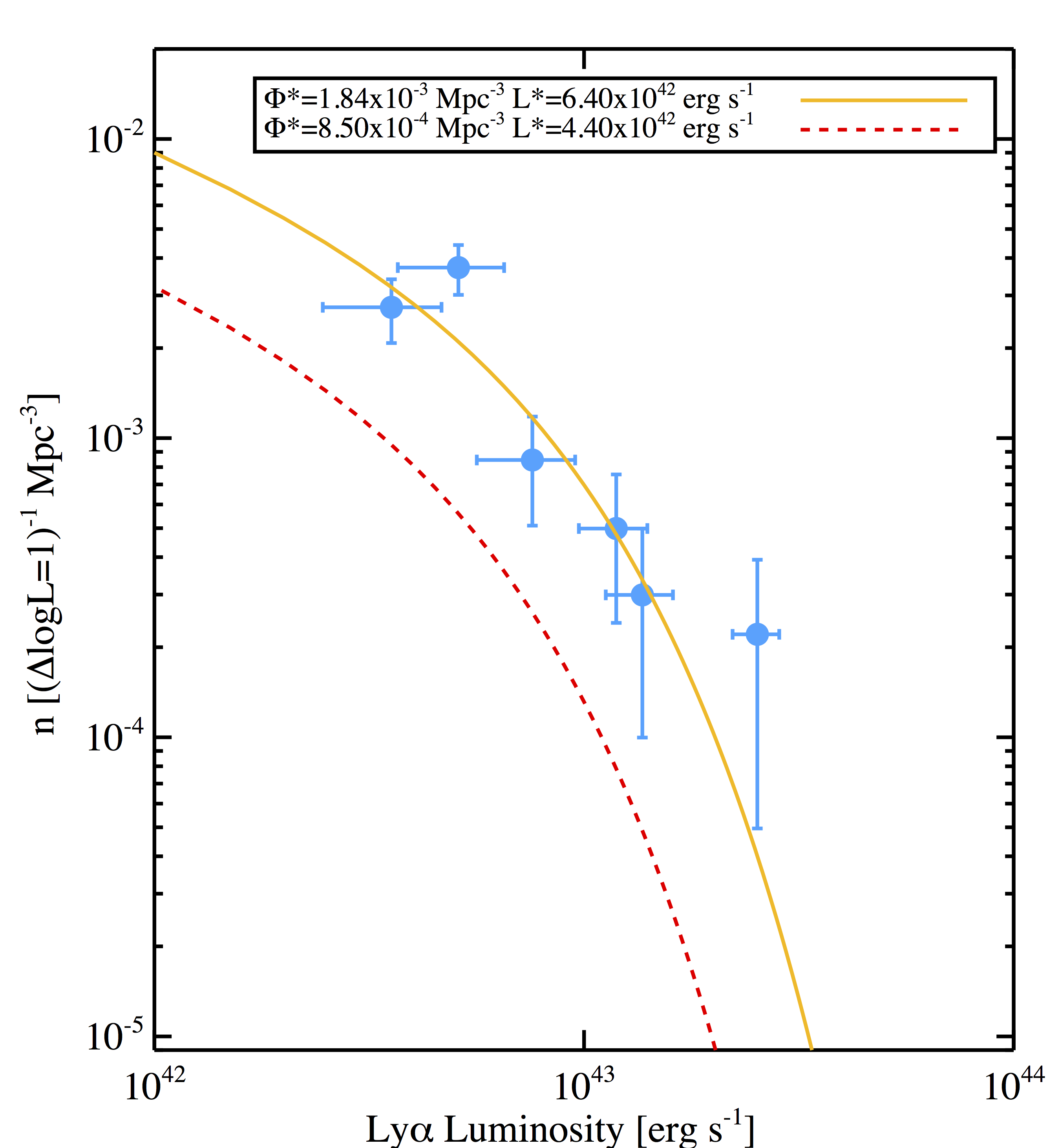}
    \caption{The $Ly\alpha$ luminosity functions of SXDS field, containing 2 massive LAEs discovered by~\citet{ouchi2010}. Our calculated number density of LAE candidates per unit luminosity per $Mpc^3$ are shown in sky-blue solid circles. Note that the number density is already taking into account the completeness-contamination correction and the success rate of spectroscopic follow-up of the photometric selected LAE candidates (i.e., $\frac{2}{3}$). The red dashed line is the best fitted luminosity function of the overall SXDS and SDF fields at redshift z=6.6~\citep{ouchi2010}. The yellow solid line is the best fitted luminosity function to our calculated LAE number densities, by performing least $\chi^2$-fitting ($\chi_{r}^2 = 1.57$). The slope of both luminosity functions are fixed at $\alpha = -1.5$; while other parameters are as indicated in the figure. Note that the $Ly\alpha$ luminosity is calculated from the SExtractor's auto-magnitudes and FWHM of F913w25 filter, assuming the LAE candidates are at z=6.5.}
    \label{fig:14}
\end{figure}

\begin{figure}
	\includegraphics[width=\columnwidth]{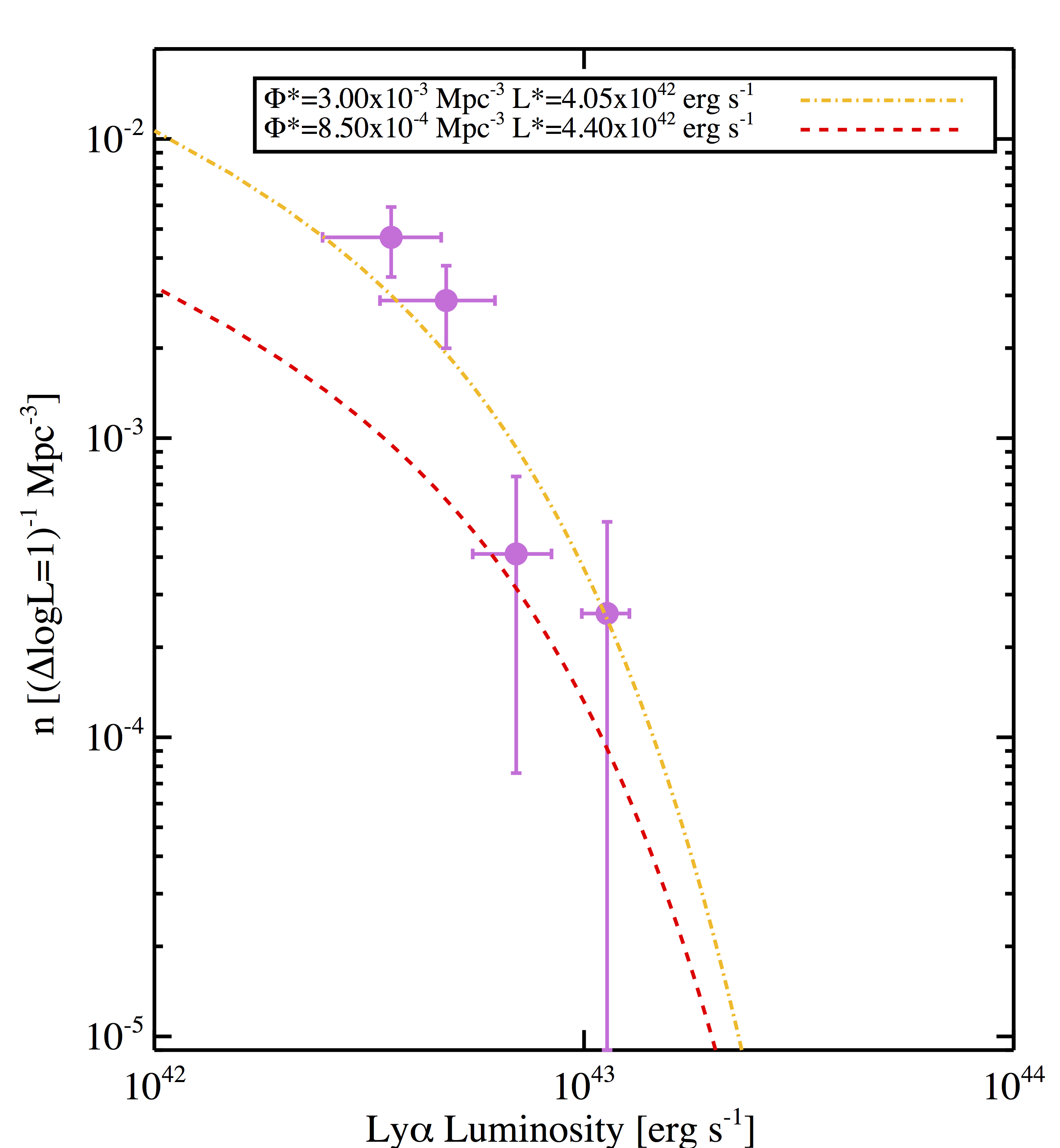}
    \caption{The $Ly\alpha$ luminosity functions of SXDS field as in Figure~\ref{fig:14}, but with only the analysis done on the left OSIRIS CCD. Our calculated number densities of LAE candidates are shown in pink solid circles. The red dashed line is the best fitted luminosity function of the overall SXDS and SDF fields at redshift z=6.6~\citep{ouchi2010}. The yellow dash-dotted line is the best fitted luminosity function to our calculated LAE number densities, by performing least $\chi^2$-fitting ($\chi_{r}^2 = 1.81$). The slope of the luminosity function is also fixed at $\alpha = -1.5$; while other parameters are as indicated in the figure.}
    \label{fig:15}
\end{figure}

\begin{table*}
	\centering
	\caption{Luminosity function parameters.}
	\label{tab:3}
	\begin{tabular}{ccccccccc} 
		\hline
		\hline
		z & $\Phi^*$ 		 		& L* 			                & $\alpha$ & $\chi_{r}^2$ &          $n^{obs}$            & $\rho_{Ly\alpha}^{obs}$              	     & $\rho_{Ly\alpha}^{tot}$                    &           \\
		   & ($10^{-3}$ Mpc$^{-3}$)  & ($10^{42}$ erg s$^{-1}$)	 & 		     & 			  & ($10^{-4}$ Mpc$^{-3}$)     & ($10^{39}$ erg s$^{-1}$ Mpc$^{-3}$) & ($10^{40}$ erg s$^{-1}$ Mpc$^{-3}$)  &    comment     \\
	     (1) & 			(2)		       & 		(3)			        &	  (4)     & 	(5)   	  &			(6)			& 		(7)			                           & 		(8)			                      & 			\\
	        \hline
		 6.5 & $1.84_{-0.61}^{+0.67}$ & $6.40_{-2.12}^{+1.55}$ &   -1.5 &      1.57         & $6.45_{-2.74}^{+2.66}$ &  $4.96_{-3.65}^{+2.99}$   & $2.09_{-0.97}^{+0.91}$     &      This work  (Full FOV)     \\
		 6.5 & $3.00_{-0.99}^{+1.87}$ & $4.05_{-1.33}^{+1.07}$ &   -1.5 &      1.81         & $10.52_{-4.47}^{+6.94}$ &        $5.12_{-3.75}^{+4.17}$        & $2.15_{-1.00}^{+1.46}$              &      This work (Left CCD)        \\		
		 6.6 & $0.85_{-0.22}^{+0.30}$ & $4.40_{-0.60}^{+0.60}$ &   -1.5 &      1.60         & $4.10_{-0.80}^{+0.90}$ &         $1.90_{-0.40}^{+0.50}$         & $0.66_{-0.08}^{+0.10}$              &     Ouchi+ 2010        \\
		\hline
	\end{tabular}
	
	\raggedright
	\textbf{Notes.} (1) Redshift; (2)-(4) best fitted Schechter parameters for  $\Phi^*$, L*, and $\alpha$ (which is fixed to -1.5); (5) reduced $\chi^2$ of the fitting function; (6, 7) observed number density and Ly$\alpha$ luminosity density calculated by integrating the best fitted Schechter function down to the observed limit of Ly$\alpha$ luminosity (i.e. $log(L_{Ly\alpha})=42.4$ erg s$^{-1}$); (8) inferred total Ly$\alpha$ luminosity density calculated by integrating the best fitted Schechter function down to $L_{Ly\alpha}=0$. 

\end{table*}

\section{Discussion}

	We can estimate the number density of z=6.5 galaxies, by computing the number density of the LAE candidates in this field. A conventional way to determine the observed number density is to integrate over the luminosity function down to our observation limit.  We can obtain the luminosity function of the LAE candidates at z=6.5 and its parameters by conducting a $\chi^2$-fitting to find the best fitted Schechter function to the binned number densities of the LAE candidates. The Schechter function used in the fitting process is expressed in Equation~\ref{eq:5}. The detailed processes in derivation of the LAE luminosity function and its best fitted parameters are discussed below.

\begin{equation}
 \Phi(L)dL = \Phi^*(L/L^*)^{\alpha}exp(-L/L^*)d(L/L^*)
	\label{eq:5}
\end{equation}

	  First, we need to calculate the expected number of LAE candidates for each magnitude bin. To do this, we use a bin size of 0.4 magnitude. This bin size ensures that there are enough candidates in each bin for the measurement to be statistically significant, while maximising the number of magnitude bins. Next, we normalise the values of the binned numbers of the LAE candidates to those with the bin size of 2.5 mag, corresponding to $\Delta log(L) = 1$. Then, we need to convert the candidates' F913 AB magnitudes into the Ly$\alpha$ luminosities. This is not a direct magnitude-flux conversion, because the F913 magnitude is the result of the combined flux of Ly$\alpha$ emission and UV-continuum. The F941 band is not deep enough to precisely determine the UV-continuum flux for all candidates. However, we can constrain the upper limit of the UV-continuum fluxes of the candidates.
	  
	  To estimate the Ly$\alpha$ luminosity for each LAE candidates, we use the expected distribution of Ly$\alpha$ equivalent width for high-z LAE as reported in many studies (e.g.,~\citet{malhotra2002},~\citet{ando2006}, and~\citet{gronwall2007}) and extrapolated to z=6.5 using the fitted relation from ~\citet{zheng2014}. We found the distribution of the rest frame equivalent width of Ly$\alpha$, $EW_0(Ly\alpha)$, at z=6.5 to be $84.7\pm 18.6 \AA$ (exponential distribution). Next, we assume that the total F913 flux comprised of Ly$\alpha$ emission and UV-continuum, $F(F913)=F(Ly\alpha)+F(UV)=f(UV)\times {FWHM(F913)} + f(UV)\times EW_{0}(Ly\alpha)*(1+z)$. The flux density of F913, then, can be expressed as $f(F913)=f(UV)\times (1+\frac{EW_{0}(Ly\alpha)\times(1+z)}{FWHM(F913)})$. The F913 flux density of an object with 25 AB magnitude is $1.34\times10^{-19} erg\:s^{-1}\:cm^{-2}\:\AA^{-1}$. Then, we can obtain F913 flux density and, as a result, $f(UV)$ of each candidates. Finally, the Ly$\alpha$ luminosity can be calculated by $L(Ly\alpha)=4\pi D_{L}^2 \times (1+z)\times EW_{0}(Ly\alpha)\times f(UV)$. The explained calculation of Ly$\alpha$ luminosity gives the L(Ly$\alpha$) of LAE-C1-01 (NB921-N-79144), which is $1.13\pm0.14\: \times 10^{43} erg\:s^{-1}$, in excellent agreement with the calculated value ($0.9\pm 1.2\: \times 10^{43} erg\:s^{-1}$) by~\citet{ouchi2010}. Nevertheless, it is worth noting that this magnitude-luminosity transformation only assumes a single emission line (i.e., Ly$\alpha$) within the F913 band. 
	  
	  The next step is to assess the possible contamination from low-z interlopers, such as [OII 3727] emitters, at z$\sim$1.4 and H$\alpha$ emitters at z$\sim$0.4, L-T dwarfs, and spurious sources as observed in the photometric catalogue of LAE candidates in SXDS and SDF fields by~\citet{ouchi2010}. First, L-T dwarfs have more or less power-law SED in our bands. As we discuss in the previous section that the adopted colour criteria already prevent against detection of objects with power-law SED. Nevertheless, we follow the treatment in~\citet{hibon2011} to asses possible number of L-T dwarfs and other interlopers within our observable window. The space density of L-T dwarfs only a few $10^{-3} pc^{-3}$~\citep{rey2010} and only the most luminous L-dwarfs can be observed up to 4kpc with our survey depths~\citep{tiney2003}. Therefore, there should be no more than 1 L-T dwarf contaminated in our candidates. Next, in order for H$\alpha$ emitters at z$\sim$0.4 to be detected in F913 and not detected in blue bands (with other emission lines, such as H$\beta$, [OII], and [OIII]) their observed EW must be at the same level or higher than our LAE candidates (i.e., $log((1+z)\times EW_0)\geq 2.8\:\AA$). We found that there is only $\sim$2\% of H$\alpha$ emitters in HST PEARS survey~\citep{straughn2009} that exhibits such a strong emission. Considering the number of H$\alpha$ at z=0.4 detected in $1\:deg^2$ survey in~\citet{geach2010} and the difference in the survey areas, we estimate the upper limit of 2 H$\alpha$ emitters as contaminants. Probably the most important source of interlopers are [OII] emitters at z=1.4; since they may not be detected in the bands blueward of F913 and appear in F913 image very much like our LAE candidates. Again, from~\citet{straughn2009}, there is $3\%$ of [OII] emitters with EW large enough to be detected in F913. With the luminosity function of high-z [OII] emitters~\citep{rig2005} and the survey volume around z=1.4 from the width of F913 band, we estimate the upper limit of 3 [OII] emitters at z=1.4 as contaminants. Therefore, we set the upper limit of 6 objects as contaminants in our catalogue of LAE candidates. This is just a little less than 13\% of the total number of the LAE candidates, and also less than the expected counting noise from 47 objects. Nevertheless, with the careful candidates selection, we want to emphasise again that all the potential interlopers should not be a major cause of contaminants in our candidates catalog. The H$\alpha$ and [OII] emitters at z=0.4 and 1.4 should have been detected in at least one of the SXDS' B, V, R, and i' bands due to their other strong emission lines (e.g. H$\beta$, [OII], and [OIII]) and strong FUV continuum, respectively. And the L-T dwarfs would not passed our colour criteria in the first place, as demonstrated in the simulation shown in Figure~\ref{fig:7}. Furthermore, we find that the spectroscopic follow-up success rate of this survey field studied by~\citet{ouchi2010} is $\frac{2}{3}$ from the total of 30 LAE candidates, with the other $\frac{1}{3}$ being interlopers and spurious. Therefore, just to be conservative, we adopt the fraction $\frac{2}{3}$ for our spectroscopic success rate as well. Then, we multiply the binned numbers of our LAE candidates by $\frac{2}{3}$, to account for the success rate. The expected number of LAEs in each bin is now normalised and accounted for the same contamination rate as in the other surveys of the same field.

	  	In addition, using the appropriate survey volume is crucial for the precise calculation of the LAE number density. In our case, we suffer from high contamination and low completeness in the 2 faintest magnitude bins (25.8-26.2 mag and 26.2-26.6 mag), especially on the right side of the OSIRIS chip as shown in Figure~\ref{fig:12} and~\ref{fig:13}. The 2 faintest magnitude bins also contain the majority of the LAE candidates. Thus, the low completeness and high contamination in these 2 bins, on the right OSIRIS chip, severely hinder our ability to detect LAE candidates in this region, causing the peculiar bimodal spatial distribution of the LAE candidates shown in Figure~\ref{fig:11}. This raises the question of whether we should use the full FOV or just the left OSIRIS chip in the calculation of the survey volume. We have first used the full FOV to calculate the survey volume, but correcting the expected number of LAEs to account for high spurious contamination and low completeness. We apply the completeness and contamination maps as in Figure~\ref{fig:12} and~\ref{fig:13} to correct for the expected number for LAEs in each bin. The process of correcting the expected number of LAEs ($N_{b}$) for completeness  and contamination is expressed in Equation~\ref{eq:6}; where, $N_{c}$, $S_{F913}(mag,x,y)$, and $C_{F913}(mag,x,y)$ are the corrected expected number of LAEs, the magnitude-spatial contamination function, and the magnitude-spatial completeness function, respectively. 

\begin{equation}
 N_{c} = N_{b}\times(1-S_{F913}(mag,x,y))/C_{F913}(mag,x,y)
     \label{eq:6}
\end{equation}

	      	  With the FWHM of the F913 filter, our survey covers the redshift range from z=6.4 to 6.6. The full FOV of the final reduced image in the F913 band is 59 arcmin$^2$. We derive a survey volume of 31,368 Mpc$^3$ (co-moving). Now, we can calculate the LAE number densities by simply dividing the corrected expected number of LAEs in each bin by the survey volume. Our LAE number densities along with the best fitted Schechter functions for our survey and the 1-$deg^2$ SXDS field~\citep{ouchi2010} are plotted in Figure~\ref{fig:14}. The error bars in number density are estimated by Poissonian statistic ($\sigma_N=\sqrt{N}$), the variation in completeness and contamination levels, and the estimated error in redshift range of the survey. The slope of the Schechter function, $\alpha$, is fixed to -1.5, as conventionally applied for many high redshift LAE observations (e.g.~\citet{mr2004},~\citet{ouchi2008, ouchi2010},~\citet{shimasaku2003, shimasaku2006}, and ~\citet{kashikawa2006}). We obtain our best fitted Schechter function via $\chi^2$-fitting with the reduced $\chi^2$ value of 1.57 ($\chi_{r}^{2}$), indicating that the fitted Schechter function represents the LAE number densities quite well. The fitted parameters from our observations and~\citet{ouchi2010} are listed in Table~\ref{tab:3}. The errors on the Schechter parameters as shown on the table are obtained from 68\% confidence level (1$\sigma$ intervals) of the $\Delta\chi_{r}^2$ distribution. 

	 To further investigate the possibility that the peculiar distribution of the LAE candidates on the right side of the OSIRIS FOV could be the result of edge effects that may affect our derived luminosity function, by repeating the analysis only on the left OSIRIS CCD (i.e., R.A.$\geq$ 2:18:20.250) where completeness and spurious corrections are the smallest. The volume corresponds to this part of the field is 48\% of the total survey volume. The constructed luminosity function of the LAE candidates on the left OSIRIS CCD are calculated and shown as pink circles in Figure~\ref{fig:15}. The LAE candidates in this part of the field have F913 AB magnitudes fainter than 25 mag. Thus, we can only construct 4 binned number densities for this group of LAE candidates, from the total of 6 magnitude bins. Nevertheless, we have found that the number densities of the 2 brightest bins are consistent with the number densities of the SXDS-N sub-field from~\citet{ouchi2010} within 1$\sigma$; while the number densities of the 2 faintest bins are well above the red dashed line by 3$\sigma$ as shown in Figure~\ref{fig:15}. The best fitted Schechter parameters to this group of LAE candidates are listed in Table~\ref{tab:3}. The observed number density of LAEs derived from the left OSIRIS CCD is consistent with the one derived for the full FOV within 1$\sigma$ as well. Thus, from the additional analysis on the luminosity function of the LAE candidates on the left OSIRIS CCD alone, we have found no conclusive evidence that all of the LAE candidates on the right OSIRIS CCD being purely spurious detections from the edge effect, nor that it could affect our measurement of the luminosity function and our conclusions. Either way, we have a strong evidence for the overdensity of LAEs at z=6.5 in this particular field.  
	  
	  The F913 medium band covers the range of redshift around z$\sim$6.4 to 6.6, while the NB921 narrow band only covers around z$\sim$6.5 to 6.6. Thus, our 3-medium-band selection should provide a longer line-of-sight depth for the high redshift LAEs survey in comparison to the traditional narrow-band selection, which yields the survey volume closer to a cubical shape in 3D. Thus, our selection method should provide a better constraint on the number density of the LAEs in this particular sub-field, even though the Ly$\alpha$ luminosity sensitivities are comparable. Considering our 3$\sigma$ limiting magnitude, this survey has the Ly$\alpha$ sensitivity down to $\sim\:log(L)=42.4$. This provides the lower limit for integration of the observed LAE number and luminosity densities,  $n^{obs}$ and $\rho_{Ly\alpha}^{obs}$. By comparing these parameters with the ones from the 1 $deg^2$ SXDS field by~\citet{ouchi2010}, we can determine the level of overdensity and also clustering signature of this sub-field. As indicated in Table~\ref{tab:3}, the overdensity level inferred from $\Phi^*$ ratio of this sub-field (this work) and the overall SXDS field~\citep{ouchi2010} is 2.16 times (3.53 times for the left OSIRIS field alone). The overdensity of the same level, leading to a protocluster around z=5.7, also has been observed(e.g.~\citet{malhotra2005}, and~\citet{wang2005}). Simulations of the clustering properties of this potential protocluster have been performed and discussed in detail by~\citet{jmre2016}, which provides further evidence that this overdensity may lead to a galaxy cluster mass $\sim10^{15}\:M_\odot$ (comparable to the Coma cluster) at $z=0$. Spectroscopic follow-up of the LAE candidates are conducted using the Multi-Object Spectrometer (MOS) capability of OSIRIS at GTC and the results will be presented in the forthcoming paper (paper-III).

\section{Conclusions}


	 We have surveyed the faint LAE population near the end of Reionisation Epoch. With our 3-band imaging approach using GTC/OSIRIS, we have successfully detected the total of 47 LAE candidates in the SXDS field containing the 2 spectroscopically confirmed massive LAEs. We have studied the level of overdensity in this particular field. While the expected number of LAEs without the overdensity would be in the order of 20 for the survey volume of $\sim30000\:Mpc^{3}$ (redshift ranges from $z=6.4-6.6$), we have found a substantially larger number of these high-z galaxies.  
	
	 After a careful analysis taking into account the spatially differential completeness and contamination correction, the success rate in spectroscopic follow-up, and the level of contamination by low-z interlopers, we have constructed the luminosity function of the LAE candidates in this sub-field and found that the best fitted parameters (i.e., $\Phi*$ and $L*$) yield the overdensity level of 2.16 times higher than the previous studies in the same field. From the clustering simulation done by~\citet{jmre2016}, this overdensity could correspond to a protocluster that would collapse and turn into a massive galaxy cluster with the total mass in the order of $10^{15}\:M_\odot$ at $z=0$, similar to the Coma cluster. However, conclusive evidence of such a protocluster in this field is yet to be confirmed via spectroscopic follow-up.

\section*{Acknowledgements}

	This work is based on the observation made during observing semester 2011B and 2012B of Gran Telescopio Canarias (GTC), which is located on Spanish Observatorio del Roque de los Muchachos of the Instituto de Astrof\'isica de Canarias, in the island of La Palma, Spain.  The observation time allocated from 14 hours via University of Florida and 20.33 hours via collaborators in Spain and Mexico. We are thankful for help and support during the observation runs from Antonio Cabrera Lavers and GTC staff. 
	
	Furthermore, this research has made use of the NASA/IPAC Extragalactic Database (NED), which is operated by the Jet Propulsion Laboratory, California Institute of Technology, under contract with the National Aeronautics and Space Administration. 
	
	We thank the funding support provided to the graduate student working on this project from the Royal Thai Government Scholarship in the section of National Astronomical Research Institute of Thailand (NARIT). JMRE, RC \& NCR (AYA2012-39168-C03-01), ESS and AM (AYA2012-39168-C03-02) and JMMH (AYA2010-21887-C04-04, AYA2010-21887-C04-02 and AYA2012-39362-C02-01) acknowledge support from the Spanish MINECO under the PNAyA grants given in parenthesis. ET, RT, and IA acknowledge support  from the Mexican Research Council (CONACYT) under grants 2008-103365, 2010-01-155046, and 2011-01-167291. 
	
	Finally, we are also extremely grateful for the referee's valuable time in reviewing this manuscript and providing thorough comments and suggestions.

\bibliographystyle{mnras}
\bibliography{bibliograph1} 
















\bsp	
\label{lastpage}
\end{document}